\documentclass[11pt,a4paper]{article}
\pdfoutput=1
\usepackage{jheppub}

\usepackage{float}
\usepackage{amsmath}
\usepackage{cleveref}
\usepackage{xcolor}
\usepackage{caption}
\usepackage{subcaption}
\usepackage{physics}
\usepackage{tikz}
\usepackage{pgfplots}
\usepackage{multirow}

\crefname{equation}{eqn.}{eqns.}
\Crefname{equation}{Eqn.}{Eqns.}
\crefname{figure}{fig.}{figs.}
\Crefname{figure}{Fig.}{Figs.}

\newcommand{\be}{\begin{eqnarray}}
\newcommand{\ee}{\end{eqnarray}}

\usepackage{mathdots}
\usepackage{arydshln}

\title{Bubble nucleation and quantum initial conditions in classical statistical simulations}

\author[a]{Anders Tranberg,}
\author[a]{Gerhard Ungersb\"ack,}

\affiliation[a]{Faculty of Science and Technology, University of Stavanger, 4036 Stavanger, Norway}

\emailAdd{anders.tranberg@uis.no}
\emailAdd{gerhard.ungersback@uis.no}

\abstract{Classical-statistical lattice simulations provide a useful approximation to out-of-equilibrium quantum field theory, but only for systems exhibiting large occupation numbers, and only for phenomena that are not intrinsically quantum mechanical in nature. In certain special circumstances, it can be appropriate to initialize such real-time simulations with quantum-like zero-point fluctuations. We will revisit these points, and investigate reports that quantum bubble nucleation rates in 1+1 dimensions can be computed through the classical evolution of such a quantum-like initial condition \cite{braden}. We find that although intriguing, the reported numerical agreement between classical-statistical simulations and the quantum nucleation rate in 1+1 dimensions is a coincidence, which is not specific to this choice of initialisation, is parameter and lattice cut-off dependent and disappears as the number of space-dimensions increases from 1+1 to 2+1.}

\begin{document}

\maketitle


\section{Introduction}
\label{sec:Intro}



\subsection{The classical-statistical approximation}
\label{sec:classical}


The classical-statistical approximation (CS) to real-time quantum field dynamics consists in replacing the evolution of the quantum operators (such as $\hat{\phi}(x,t)$) by classical dynamics of an ensemble of random initial conditions. The ensemble is taken to reproduce the initial correlators of the quantum system, and each random member of the ensemble is evolved by means of the classical equations of motion. The expectation values of observables are then computed as averages over the ensemble. 

The CS approximation is reliable only when the occupation numbers (particle numbers) $n_{\bf k}$ of the fields are large, $n_{\bf k}\gg 1$ (se for instance \cite{Hertzberg:2016tal}). For a typical massive scalar field, the field equation reads
\begin{eqnarray}
(\partial_t^2-\partial^2_x+m^2)\hat{\phi}(x,t)=-\frac{dV_{\rm nl}(\hat{\phi})}{d\hat{\phi}}(x,t).
\end{eqnarray}
with $V_{\rm nl}$ denoting non-linear self-interactions.
When non-linearities are small, individual momentum modes behave as harmonic oscillators, and we may write  for the field and canonical momentum operators
\be
\hat{\phi}(x,t)=\int\frac{d^dk}{(2\pi)^d}\hat{\phi}_{\bf k}(t) e^{i{\bf kx}},\quad
\hat{\phi}_{\bf k}(t) = \hat{a}_{\bf k} f_{\bf k}(t) + \hat{a}^\dagger_{\bf k} f_{\bf k}^*(t),\\
\hat{\pi}(x,t)=\int\frac{d^dk}{(2\pi)^d}\hat{\pi}_{\bf k}(t) e^{i{\bf kx}},\quad
\hat{\pi}_{\bf k}(t) = \hat{a}_{\bf k} \dot{f}_{\bf k}(t) + \hat{a}^\dagger_{\bf k} \dot{f}_{\bf k}^*(t),
\ee 
Then the occupation number of a field momentum mode is given by
\be
\langle\phi_{\bf k}(t)\phi_{\bf k}^\dagger(t)\rangle=\frac{n_{ \bf k}+\frac{1}{2}}{\omega_k}= \frac{\langle a^\dagger_{\bf k}a_{\bf k}+a_{\bf k}a^\dagger_{\bf k}\rangle}{2}|f_{\bf k}(t)|^2= 
\left(\langle a^\dagger_{\bf k}a_{\bf k}\rangle+ \frac{1}{2}\right)|f_{\bf k}(t)|^2.
\ee
where by standard convention, we have taken $|f_{\bf k}|^2(0)=1/\omega_{\bf k}$, $\omega_{\bf k}^2=\bold{k}^2+m^2$.
The zero-point fluctuations (corresponding to the zero point energy of a harmonic oscillator) is the ``1/2'', while the excitations of the system above the vacuum are the $n_{\bf k}$. The classical limit corresponds to $n_{\bf k}\gg \frac{1}{2}$. 

This argument relies on the particle numbers $n_{\bf k}$ which is a quasi-particle concept, valid at weak coupling. The argument may be generalised and made more precise in the context of the Keldysh formalism and Kadanoff-Baym equations for the real-time correlation functions (see for instance \cite{Aarts:1997kp}, and a related discussion in \cite{Hertzberg:2016tal}).
It is convenient to introduce the ``statistical'' and ``spectral'' propagators 
\be
F(x,y) = \frac{1}{2} \langle [\phi(x),\phi(y)]_+ \rangle,\quad \rho(x,y) = i \langle [\phi(x),\phi(y)]_- \rangle, 
\ee
so that the complete propagator (on the Keldysh contour $\mathcal{C}$) may be written:
\be
G(x,y) = \langle T\{\phi(x),\phi(y)\}\rangle= F(x,y)-\frac{i}{2}\textrm{sign}_{\mathcal{C}}(x^0-y^0)\rho(x,y).
\ee
The real-time evolution of correlators may be expressed through diagram expansions in terms of $F$ and $\rho$, both for quantum and classical field theory \cite{Aarts:1997kp,Aarts:2001yn} (see also \cite{Rajantie:2006gy} for an explicit application and comparison). In the classical approximation, certain diagrams turn out to be absent\footnote{In terms of the Keldysh field basis of $\phi_{cl}$ and $\phi_q$, for instance in $\lambda \phi^4$-theory, the 3-$\phi_{q}$ vertex is absent, and any diagram involving this vertex.} so that whenever the quantum theory self-energy contains a combination of the form\footnote{For different theories, diagrams and self-energies, prefactors may vary. For instance, in $\lambda\phi^4$-theory, the sunset diagram produces $3F^2-\rho^2/4$ in the self-energy component for $\rho$, while $F^2-3\rho^2/4$ appears in the self-energy component for $F$ \cite{Arrizabalaga:2005tf}.}
\be
\Sigma_{\rm quantum}\simeq F^2(x,y)-\rho^2(x,y)/4,
\ee
in the classical theory the same diagram has only 
\be
\Sigma_{\rm classical}\simeq F^2(x,y).
\ee
Hence, the classical approximation is good, whenever $\rho^2$ can in fact be neglected, $F^2\gg \rho^2$. For weak coupling and in the quasi-particle picture, $F\simeq (n_{\bf k}+1/2)/\omega_{\bf k}$, $\rho\simeq 1$, in which case the criterion for classicality again amounts to $n_{\bf k} \gg 1$. 

A third fully non-perturbative derivation of the classical-statistical approximation follows directly from the Keldysh contour path integral \cite{Mou:2019gyl}\footnote{The primary focus of \cite{Mou:2019gyl} is the subsequent evaluation of the path integral in the Picard-Lefschetz formalism, but the relation to the CS approximation is independent of that further application.}. In short, whereas the quantum result comes about from averaging over the field variables  at all times on the Keldysh contour (all ``paths''), the CS approximation amounts to only averaging over the field variables at the initial time, corresponding to the ensemble of initial conditions. 

The authors of \cite{Mou:2019gyl} proceed to show that the phenomenon of tunneling in quantum mechanics may be computed from the complete path integral or the Schr\"odinger equation yielding the same result, while the CS approximation fails to correctly reproduce the tunneling rate. Similarly, the CS approximation fails to describe the famous quantum violation of Bell or Leggett-Garg inequalities \cite{Millington:2020vkg}, even in the free-field limit. In fact, the CS approximation may only be used to compute certain ``classical'' observables.\footnote{The ones involving the $\phi_{\rm cl}$-field and not the $\phi_q$-field in the Keldysh basis.} One obvious example is that the fundamental commutator $[\phi(x),\pi(y)]_-=i\delta^d(x-y)$, which is inherently ``quantum'', vanishes in the CS approximation.\footnote{Note that in actual classical field theory, one may define objects with similar properties, such as the Poisson bracket of canonical variables. But although much of the concrete numerical computation is the same, conceptually classical field theory and the CS approximation to quantum fields are distinct. } The failure of the CS approximation to reproduce the tunneling rate in quantum mechanics (as given by solving the Schr\"odinger equation) is a standard (see \cite{hertz} for a recent analysis in the present context). 

\subsection{Classical-Statistical simulations and the ``half''}
\label{sec:classtat}

There is nothing to prevent us from performing CS computations from any initial condition, provided we are able to somehow generate the configurations making up the initial ensemble. One example is the classical thermal equilibrium-like state, parametrized by some temperature $T$, $n_{\bf k}+1/2=T/\omega_k$, which up to corrections from non-linear interactions is a fixed point of the dynamics. But evolving some generic initial ensemble amounts to classical field theory from a non-equilibrium initial state, not necessarily with any connection to a quantum system.

As we have seen, only for large occupation numbers (or large $F$) can a CS computation be expected to yield a good approximation to the quantum result, and only for appropriate observables. Also, the approximation can in many cases only be expected to hold for a finite time, since the late time asymptotic state is the classical equilibrium\footnote{Which is badly defined in the continuum limit, but has meaning on a finite lattice with a momentum cutoff.} rather than the quantum equilibrium.

Most physical phenomena are dominated by some characteristic momentum range, and the spectrum of momentum modes split up into regions with large ($\gg 1$), moderate ($\simeq 1$) and small ($<1$) occupation numbers. As long as the modes relevant for the phenomenon of interest are highly occupied, the expectation is that classical dynamics will give a reliable result when applied to all modes. Typical examples include (near-to-)equilibrium systems at high temperature (see for instance \cite{Moore:1999fs,Berges:2007re}), large objects such as topological defects or sphalerons (for instance \cite{Rajantie:2010tb,DOnofrio:2014rug}), as well as high-occupancy phenomena such as resonances and instabilities (for instance \cite{Rajantie:2000nj,Greene:1997fu,Kofman:1997yn,Felder:2000hj,Bodeker:2007fw,Rebhan:2004ur,Tranberg:2003gi,Garcia-Bellido:1999xos}). 

For a few very specific cases, a very special initial condition has been employed dubbed ``the quantum half'' \cite{Rajantie:2000nj,Garcia-Bellido:2002fsq,Smit:2002yg}. The prescription is to represent an initial quantum vacuum state with $n_{\bf k}+\frac{1}{2} = \frac{1}{2}$ by an ensemble of classical initial conditions, and evolve the system classically from there. In most cases, this is very problematic, since $n_{\bf k}\gg 1$ is certainly not satisfied, and the energy density of the initial state is cut-off dependent and divergent. 

Another issue is that while the true quantum dynamics ensures that the zero-point fluctuations stay put in each mode \cite{Arrizabalaga:2004iw}, allowing only the exchange of the $n_k$ between modes, the classical dynamics does not distinguish between the $n_{\bf k}$ and the $1/2$ excitations, and will allow all to be exchanged. Extracting energy from the zero-point fluctuations in this way is an unphysical effect, which is negligible if $n_{\bf k}+1/2$ is anyway large, but may be very important when $n_{\bf k}+1/2\simeq 1/2$.

However, one property can make it reasonable to describe the quantum dynamics of a quantum-like ``half'' initial condition by the CS approximation: For non-interacting fields, the operator field equations are linear, as described above allowing us to expand the Heisenberg field operators as independent time-independent harmonic oscillators. To compute numerically
\be
\langle\phi_{\bf k}(t)\phi_{\bf k}^\dagger(t)\rangle = \left(a_{\bf k}^\dagger a_{\bf k}+\frac{1}{2}\right) |f_{\bf k}(t)|^2,
\ee
we only need to solve for $f_{\bf k}(t)$, while the $a_{\bf k}$, $a_{\bf k}^\dagger$ are time-independent operators containing the information about the initial state. Since the evolution is linear, it makes no difference whether we evolve from the initial condition $f_{\bf k}(0)=1/\sqrt{\omega_k}$ and multiply by $n_{\bf k}+1/2$ at the end, or whether we classically evolve an ensemble of initial conditions $\phi_{\bf k}(0)$ with the property that $\langle \phi_{\bf k}(0)\phi^\dagger_{\bf k}(0) \rangle = (n_{\bf k}+\frac{1}{2})/\omega_k$. This is the CS approximation, and so for a non-interacting field, the approximation to the evolution is exact, irrespective of $n_{\bf k}$\footnote{See \cite{Millington:2020vkg} for a detailed discussion of what observables this prescription allows us to compute.}.

This means that for systems, where for some reason the occupation numbers grow large while still in the linear regime (for small coupling, say), we are allowed to initialise the classical system in the quantum-vacuum like state $n_{\bf k}=1/2$, and evolve the system using classical equations of motion throughout; at early times because the system is linear, at late times because the system has large occupation numbers. We only require that occupation numbers grow large before self-interactions become important.

To summarize: The ideal prescription to simulate a phenomenon arising from a quantum vacuum initial condition is to 1) start off with $1/2$ in all modes, 2) evolve them all with the (quantum, but equivalently classical) linear equations until non-linear self-interactions become important, 3) discard all the modes that have not by then acquired large occupation numbers, and only 4) continue the now classical evolution of the highly occupied modes. Various levels of adherence to these rules can be argued for on a case-by-case level. 
Examples, where this applies include: 
\begin{itemize}
\item The primordial perturbations responsible for cosmological structure formation. These are the zero-point fluctuations of a weakly coupled scalar field, that grow because the accelerated expansion of space introduces non-adiabatic evolution of the modes \cite{Mukhanov:1990me}. This is one instance of the phenomenon known as ``squeezing'' of an initial vacuum state. Observations show that non-Gaussianities are minute, and so the entire early-time evolution from vacuum fluctuations ($1/2$) to non-vacuum ($n_{\bf k}\gg 1$) may be simulated using (almost linear) classical evolution. 
\item Resonant preheating after inflation arises when at the end of inflation, the oscillating inflaton mean field is in resonance with certain field modes (whether of another field or the inflaton itself). This resonance amplifies these modes from an initial vacuum state to large occupation numbers \cite{Greene:1997fu,Kofman:1997yn} (for a recent review, see \cite{Amin:2014eta}). Since the self-interaction is usually quite small ($\lambda\simeq 10^{-12}$ or smaller for many inflation models), occupation numbers can grow very large before non-linearities become important. And so the CS approximation is valid all the way from the quantum vacuum initial state. 
\item  Tachyonic preheating (or spinodal decomposition) occurs in hybrid inflation-type models, where a negative curvature of the potential $V$ triggers an instability of certain modes $k^2+V''<0$ \cite{Felder:2000hj}. These modes grow exponentially, until self-interactions become important. If the self-interaction is small, the classical evolution again holds from the initial quantum vacuum state (when the evolution is linear), and also in the subsequent non-linear regime, because occupation numbers are by then $\gg 1$ (see for instance \cite{Garcia-Bellido:2002fsq,Smit:2002yg}). 
\item Certain plasma instabilites in gauge theories can also be described as unstable modes, at weak coupling \cite{Bodeker:2007fw,Rebhan:2004ur}. As these acquire large occupation numbers, the CS approximation can be applied also in the context of the approach to thermal equilibrium in heavy-ion collisions. 
\end{itemize}
A final point worth mentioning is that the classical regime with large occupation numbers does not imply that one particular classical realization (one member of the ensemble) is singled out. All observables must be computed as statistical expectation values over the whole classical ensemble of configurations, which is then expected to reproduce well the expectation values over the wave function (or density matrix) of the quantum system.

\subsection{Classical simulations of vacuum decay}
\label{sec:intitcond}

A quantum system at zero temperature in a local potential minimum (a ``false'' vacuum) may decay into a state in the global minimum (the ``true'' vacuum) through quantum mechanical tunneling. In the Euclidean formulation of quantum field theory the transition is described by an instanton \cite{colman1}, and from a path integral point of view, the transition is mediated by non-classical paths, paths that do not satisfy the classical equations of motion. The transition rate is straightforwardly computed in quantum mechanics, but is substantially harder to extract in quantum field theory.

In \cite{braden}, an approximate agreement was reported between the instanton computation of the transition rate in 1+1 space-time dimensions, and the CS evolution of a vacuum (``half'') initial state in the unstable vacuum. The authors were surprised and intrigued by their result, since tunneling is precisely the type of very quantum processes, where one would expect the CS approximation to fail. Indeed in quantum mechanics (field theory in 0+1 dimensions), the CS approximation does fail to reproduce the quantum tunneling rate \cite{Mou:2019gyl}. 

Classical simulations of bubble nucleation are natural in the context of a finite-temperature phase transition, where the initial state is described by the finite temperature distribution of occupation numbers above the unstable vacuum. Then the transition is a classical effect whereby there is some (Boltzmann) probability that the ambient thermal fluctuations manage to spontaneously form a true-vacuum bubble, large enough to make it over the potential barrier and expand to eventually fill the whole of space (we will return to this point in more detail below). 
It follows that the finite-temperature bubble nucleation rate can in principle be computed by classically evolving all field configurations starting in the local potential minimum, and then averaging them over the initial Boltzmann distribution, schematically
\be
\label{eq:clasrateaverage}
\Gamma_{\rm Finite\,T} = \int P_{\rm Boltzmann,T}[\textrm{configuration}]\times \textrm{transition rate of the configuration}.\nonumber\\
\ee
The result of \cite{braden} would suggest that the quantum tunneling rate follows from the same set of classical trajectories, but averaged over the quantum vacuum-like initial distribution
\be
\Gamma_{\rm Quantum} = \int P_\textrm{Vacuum, $\frac{1}{2}$}[\textrm{configuration}]\times \textrm{transition rate of the configuration}.\nonumber\\
\ee
This is a surprising result, and warrants further scrutiny. In particular, since classical evolution conserves energy, it would imply that quantum tunneling is simply the classical evolution of the subset of the initial condition ensemble, that have enough energy to nucleate a bubble. 

In \cite{hertz}, the numerical computations of \cite{braden} were reproduced, although crucially it was pointed out that to get the reported agreement between numerical and instanton results, a ``fudge'' factor $\epsilon$ had to be introduced. The agreement occurs for $\epsilon\simeq 1/2$ which amounts to rescaling the zero-point fluctuations from $n_{\bf k}=\frac{1}{2}$ to $\frac{1}{8}$. The authors of \cite{hertz} then carried out similar simulations of different initial conditions, and different models as well as an analysis related to cut-off dependence and renormalisation. The conclusion remained, that only when rescaling the amplitude of the initial conditions by an arbitrary factor $< 1$ is it possible to find the approximate agreement reported in \cite{braden}. 

We will expand further on that analysis, and show that the reported agreement is indeed a coincidence to do with the choice of the parameters of the model, the lattice cut-off and the fudge factor, and that it is not specific to the ``half'' initial condition. We will also generalise the simulations to 2+1 dimensions, and show that there is no agreement there. We will see that there are some essential differences between nucleation in 1+1 and higher dimensions.

\section{Tunneling and Bubble Nucleation}
\label{sec:tunneling}

Consider a potential $V$ with two non-degenerate minima, with a barrier in-between. If the system is initially in the local minimum with highest energy, a transition may occur whereby the system moves to the global minimum with lowest energy (``false vacuum decay''). Energetically, it is very expensive for the field to move across the barrier in all of space simultaneously. Instead, one local region of space (a bubble) is created with the field in the global vacuum inside, in the local minimum outside, and with the field continuously interpolating between the two on the boundary (the wall).

\subsection{Classical Bubble Nucleation}
\label{sec:clastunneling}

Classical bubble nucleation is the process by which random classical fluctuations (for instance in equilibrium at a temperature $T$) by chance organise themselves into such a bubble. This happens all the time, but most bubbles are so small, that they collapse again. The energy criterion controlling the process is the balance between the energy cost of creating the bubble wall, interpolating between vacua, and the energy gain from the inside of the bubble having a lower potential energy than when the bubble is not there. In the simplest approximation one may write
\be
E = \textrm{Surface}\times\sigma   +\textrm{Volume}\times\Delta V  ,
\ee
where $\sigma$ is the surface tension, the energy associated with the interpolating field wall, and $\Delta V$ is the difference in potential at the two minima $V_{\rm global}-V_{\rm local}$ (which is negative). 
In 1+1 dimensions, the volume is the distance between walls, $2R$, while the surface is just a factor of 2 (2 walls),
\be
E_{1}= 2\sigma +  2R\Delta V,
\ee
In order for a transition to happen, a random fluctuation has to occur that creates a pair of walls. Once these walls are established, there is no further energy cost in increasing the size of the bubble. The total energy is linearly decreasing with increasing $R$. We define the critical energy and the critical radius to be
\be
E_{\rm crit,1 }= 2\sigma,\qquad R_{\rm crit} = 0\quad(\textrm{or the width of a wall}).
\ee
In 2+1 dimension, things are qualitatively different. Now
\be
E_{2}= 2\pi R\sigma + \pi R^2\Delta V ,
\ee
which is maximised to give the saddle point solution 
\be
E_{\rm crit, 2}=\frac{\pi \sigma^2}{\Delta V} ,\qquad R_{\rm crit, 2}=-\frac{\sigma}{\Delta V}.
\ee
In most cases, a random fluctuation does not acquire this critical radius, and the transition does not complete. The bubble shrinks again. But occasionally, a critical-size bubble is generated, which then continues to grow.
In 3+1 dimensions, we have
\be
E_{3}= 4\pi R^2\sigma + \frac{4\pi}{3} R^3\Delta V ,
\ee
so that
\be
E_{\rm crit, 3}=\frac{16\pi}{3}\frac{\sigma^3}{\Delta V^2} ,\qquad R_{\rm crit, 3}=-\frac{2\sigma}{\Delta V}.
\ee
Throughout, we have assumed that the bubble is spherical, since this maximises the volume/area. There will be subleading contributions from many other near-spherical configurations. 

In thermal equilibrium, the bubble nucleation rate is then proportional to the Boltzmann probability of a large enough random fluctuation
\be
\frac{\Gamma}{V t}\propto e^{-E_{\rm crit}/T}.
\ee
Dividing by the volume $V$ (not to be confused with the potential) and $t$ normalises the rate to unit volume and time, respectively. A more detailed numerical analysis along the lines of (\ref{eq:clasrateaverage}) allows the direct computation of this quantity \cite{Moore:2000jw,Moore:2001vf,Gould:2022ran}.

In a non-thermal environment, for instance a state with some non-thermal occupation numbers $n_{\bf k}$, the probability of creating such bubbles will depend on the state. As for the thermal equilibrium state, it may require that random multi-wavelength fluctuations manage to organise themselves into a large enough bubble configuration. But one could also imagine a state with only long-wavelength fluctuations (say, of size $R_{\rm crit}$), in which critical-size bubbles are ubiquitous. 

There is also the possibility that the state (whether thermal or not), simply has an energy density (much) larger than the height of the potential barrier. Then the system hardly notices, that the minima are separated, and will not need to minimise energy into a spherical bubble to perform the transition. In this case, transitions are common and fast. If the energy density is larger than $|\Delta V|$, one may also get transitions back again. 

Finally, there is the possibility that the entire physical volume has too little energy to even make a single critical bubble. This is only a practical issue in a numerical simulation of a finite volume, and hence finite total energy. Then a transition will never happen, if the dynamics are classical and energy conserving. 

\subsection{Quantum Bubble Nucleation}
\label{sec:quantunneling}

Quantum tunneling is most apparent in situations where a barrier separates two local minima of the potential, and the energy of the state is smaller than the height of the barrier. In quantum mechanics (field theory in 0+1 dimensions), starting in one minimum, one may straightforwardly solve for the wavefunction of the system, giving a non-zero probability of finding the particle inside, and on the other side of the barrier. In time, there is an ever increasing probability for the particle to be measured in the other minimum. In the case when the second minimum is in fact the global minimum, we speak of vacuum decay.

In field theory, the analogous process can also be interpreted in terms of Euclidean instanton paths, famously in \cite{colman1}. This instanton is a 4-D spherically symmetric saddle point of the Euclidean action. One may again write down
\be
S_{\rm crit, 4}=\frac{27 \pi^2}{2}\frac{\sigma^4}{\Delta V^3} ,\qquad R_{\rm crit, 4}=-\frac{3\sigma}{\Delta V}.
\ee
To a good approximation, the rate of tunneling may then be written as
\be
\frac{\Gamma}{Vt} \propto e^{-S_{\rm crit,4}},
\ee
but keeping in mind that this is the saddle point action rather than an energy, and that no temperature is involved. 

\subsection{The wall tension $\sigma$}
\label{sec:wealltension}

Whereas $\Delta V$ is simply the difference between potential minima, computing the wall tension $\sigma$ in the general case requires knowledge of the wall profile. For classical nucleation, an approximation is found by solving the (spherically symmetric) equation of motion for a static field profile interpolating between the two minima:
\be
\label{eq:bounce}
\partial_t\phi=0\rightarrow \Big(\partial_r^2+\frac{(d-1)}{r}\partial_r \Big)\phi =\frac{dV}{d\phi},
\ee
with the boundary conditions, $\phi(r=\infty)=\phi_{\rm local}$, $\partial_r\phi(0)=0$, $\phi(0)=\phi_{\rm global}$.

For $d=1$, this looks like time evolution in the potential $-V$, and is usually solved by numerical
means (shooting) \cite{CosmoT}. Then one may compute the wall tension as
\be
R^d\sigma = \int_0^{\infty} dr\, r^{d}\Big[\frac{1}{2}(\partial_r\phi)^2+V(\phi)\Big].
\label{eq:anatension}
\ee
In the limit when the wall is much thinner than the size of the bubble, the term
$(d-1)/r$ may be neglected. Then it is not necessary to know
the detailed shape of the wall, as one may rewrite (\ref{eq:anatension}) into
\be
\sigma = \int_{\phi_{\rm local}}^{\phi_{\rm global}} \sqrt{2V(\phi)}d\phi
\ee
which is easily computed, at least numerically.

For the 4-dimensional instanton we must first rotate to Euclidean space, the saddle point equation in $d+1$ dimensions becomes
\be
(\partial_\tau^2+\partial_{\bf x}^2)\phi=\frac{\partial V}{\partial \phi},
\ee
which in 4-dimensional spherical coordinates is equivalent to Eq. (\ref{eq:bounce}), in one dimension higher. Hence for a thin wall, the calculation of the wall tension proceeds in exactly the same way. This does not directly imply a relation between the tunneling rates, since $E_{\rm crit}$ and $S_E$ are very different objects.


\subsection{A convenient toy model potential}
\label{sec:potential}


\begin{figure}[ht]
  \centering
  \includegraphics[width=0.7\textwidth]{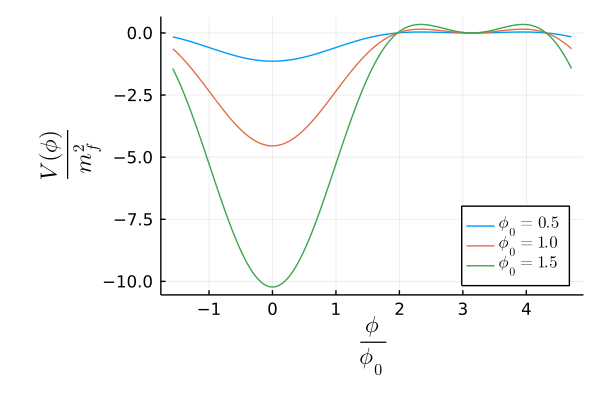}
  \caption{The potential in (\ref{eq:action}) for different values $\phi_0$.}
  \label{fig:Potential}
\end{figure}

Following \cite{braden} we will focus on a specific potential, defined by the action
\be
S = \int dx^{d+1} \left[\frac{1}{2} \partial_{\mu} \phi \partial^{\mu} \phi - V_0 \Big( -\cos \bigg( \frac{\phi}{\phi_0} \bigg) + \frac{\lambda^2}{2} \sin^2 \bigg( \frac{\phi}{\phi_0} \bigg) -1 \Big)\right].
\label{eq:action}
\ee
It is parameterized by three quantities, $\lambda$, $\phi_0$ and $V_0$. 
For $\lambda>1$ the periodic potential has global and local minima at $\phi=2n \pi\phi_0$ and $\phi=(2n+1)\pi\phi_0$, respectively, with integer $n$. The potential is chosen to have $V(\phi_{\rm local})=0$ and $\Delta V=-2V_0$, and we define the masses
\be
m_{f}^2 &=& \frac{d^2V}{d\phi^2} \Big|_{\phi=\phi_{\rm local}} = \frac{V_0}{\phi_0^2} (-1 + \lambda^2),\\
m_{t}^2 &=& \frac{d^2V}{d\phi^2} \Big|_{\phi=\phi_{\rm global}} = \frac{V_0}{\phi_0^2} (1 + \lambda^2).
\ee
The height of the potential barrier separating the two minima is given by
\be
V_{\rm max} = m_f^2 \phi_0^2 \Big( \frac{-1 + \lambda^2}{ 2 \lambda^2} \Big).
\ee
We will follow \cite{braden} and set $\lambda=1.2$. The potential is therefore parametrized by $m_f$ and $\phi_0$. In this parametrization $\phi_0$ fixes the location of the local vacuum but also influences the relative height of the potential barrier. We show in Fig.~\ref{fig:Potential} the potential for example sets of parameters. We will compute the bubble nucleation rate primarily as a function of $\phi_0$, and from the potential alone, we expect the rate to decrease with increasing $\phi_0$.


\subsection{Numerical implementation}
\label{sec:numerical_2}


\begin{figure}[ht]
  \centering
  \includegraphics[width=0.7\textwidth]{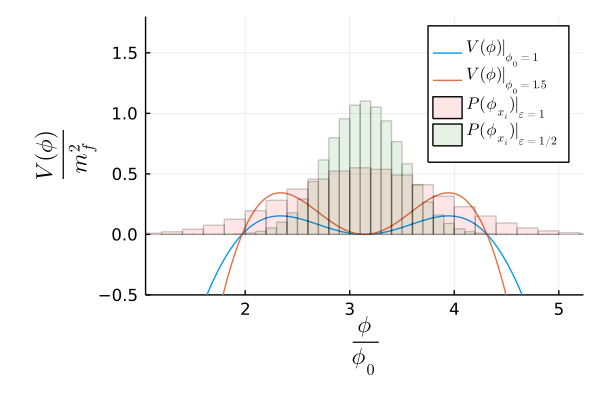}
  \caption{The potential for $\phi_0=1$ and $\phi_0=1.5$ for $am_f=0.8$. 
  The superimposed histograms show the initial distribution of $\phi(x)$ which depends on both the fudge factor $\epsilon$ and the mass $am_f$.}
  \label{fig:Potdist}
\end{figure}

We discretize the action on a space-time lattice, and solve the classical equation of motion,
\begin{align}
\dot{\phi} &= \pi, \\
\dot{\pi} &= \nabla^2 \phi - V^{\prime}(\phi).
\end{align}
A symplectic integrator scheme is used to ensure energy conservation for long simulation times.

The lattice has periodic boundary conditions and the number of lattice sites per dimension and the spacing are denoted as $N_x$ and $a$, giving the linear lattice size $L=N_xa$. We recast the lattice action in lattice units, whereby all dimensionfull quantities appear in dimensionless versions by means of the lattice spacing as $am_f$, $a^{d+1}V_0$, $a^2k^2$ and so on. Consequently, the dispersion relation on the lattice is determined by the discretized Laplacian and given by
\be
a^2\omega_k^2 = k_L^2 + a^2m_f^2, \qquad k_L^2 = \sum_{i=1}^d 2 - 2 \cos(k_i),
\ee
where for each spatial dimension $i$, $k_i=n_i \frac{2 \pi}{N_x}$ for $n_i =-N_x/2+1, ...,N_x/2$.
The quantity $am_f$ then defines the lattice cut-off, since if the maximum momentum is $a\Lambda\simeq \pi $ then the cut-off in physical units is $\Lambda/m_f=\frac{\pi}{am_f}$. As $am_f$ decreases, the cut-off increases. We will in the following only explicitly write out powers of $a$ when needed. 

The quantum-like initial conditions are Gaussian distributed field fluctuations defined by 
\be
\label{eq:fluct}
 \langle \phi_{\bold{k}} \phi_{\bold{k^{\prime}}} \rangle &= \epsilon^2 \frac{1}{2 \omega_k} \delta^d _{\bold{k} - \bold{k^{\prime} }} \qquad
 \langle \pi_{\bold{k}} \pi_{\bold{k^{\prime}}} \rangle &= \epsilon^2 \frac{\omega_k}{2} \delta^d _{\bold{k} - \bold{k^{\prime}}}
\ee
These vacuum fluctuations are added to a homogeneous field placed initially at the local minimum 
$\phi(x)= \pi \phi_0$. 

The ``fudge factor'' $\epsilon$ was introduced by \cite{hertz} to parametrically fit
tunneling rates to instanton results. $\epsilon =1 $ is the physical value that mimics a quantum vacuum state, whereas other values have no obvious physical interpretation. As we will see, and consistent with \cite{hertz}, the apparent agreement between CS results and the instanton rate arises for $\epsilon\simeq 0.5$.

\begin{figure}[!hb]
    \centering
    \includegraphics[width=8cm]{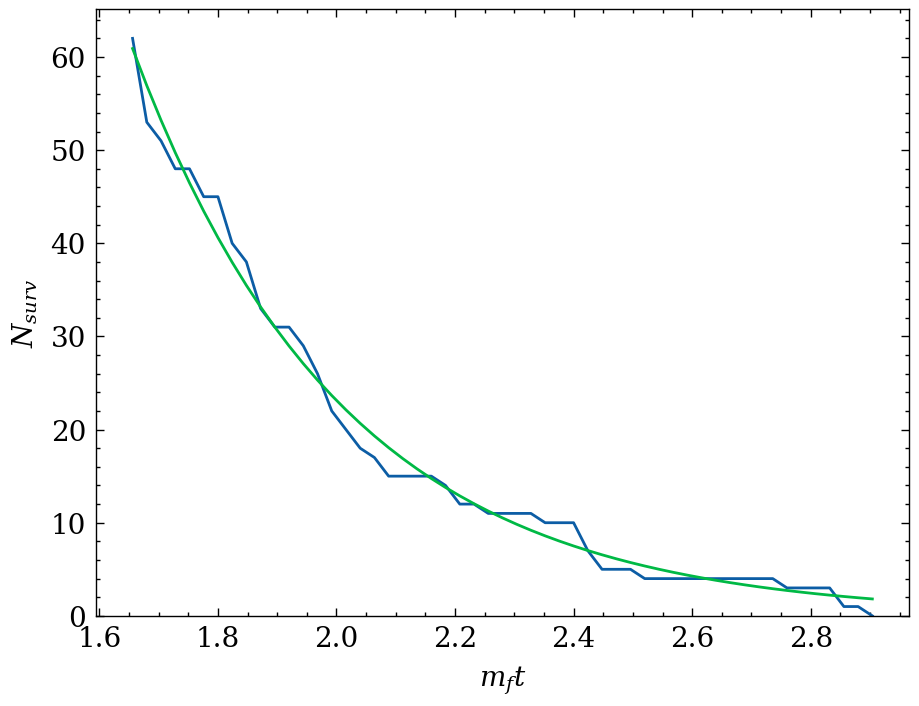}
    \caption{Example in 1+1 dimensions of the time dependence $N_{\rm surv}(t)$ and a fit to the form $N_0 e^{-t \Gamma}$. The plot shows the exponential behaviour starting from roughly 60 percent of the total number of configurations, which was $N=100$.}
    \label{fig:1+1fitexample}
\end{figure}

In preparation of the later discussion, it is instructive to generate a single initial condition $\phi(x)$, and simply compute the distribution of local field values. Fig.~\ref{fig:Potdist} shows a histogram superimposed on the potential. For a fudge factor of $\epsilon=0.5$, we see that the entire field configuration is inside the false vacuum initially. But for $\epsilon=1$, already at the initial time, the field is on the other side of the potential barrier in some small parts of space. 

Following \cite{braden,hertz}, as the simulation proceeds, we monitor the observable $\langle\cos(\phi/\phi_0)\rangle$, where $\langle . \rangle$ refers to the ensemble average, to define whether a configuration has transitioned to one of the neighboring global minima. 
For homogeneous configurations at the local/global minima this observable takes the value $-1$ or $+1$.
A configuration is then said to have transitioned if 
\be
\langle\cos(\phi/\phi_0)\rangle > \langle\cos(\phi/\phi_0)\rangle_{t=0} + 10 \Delta_{t=0} , 
\ee
where $\Delta_{t=0}$ is the standard deviation of the same observable $\cos(\phi/\phi_0)$ at the initial time. 

Given an ensemble of $N$ configurations, we define $N_{\rm surv}(t)$ to be the number of these configurations that by a given time $t$ have not yet transitioned. We then perform a fit to the form
\be
N_{\rm surv}(t) = N_0 e^{-\Gamma t},
\ee
where $N_0$ refers to the starting point of the fit. Then $\Gamma$ is the bubble nucleation rate. 
Typically it takes some time before the configurations begin to transition. The fit was therefore performed from a time when $N_0$ was 60 percent of the total configurations in the simulation. An example is shown in Fig. \ref{fig:1+1fitexample}. 


\section{The rate in 1+1 dimensions.}
\label{sec:1D}


\begin{figure}[h]
    \centering
    \includegraphics[width=12cm]{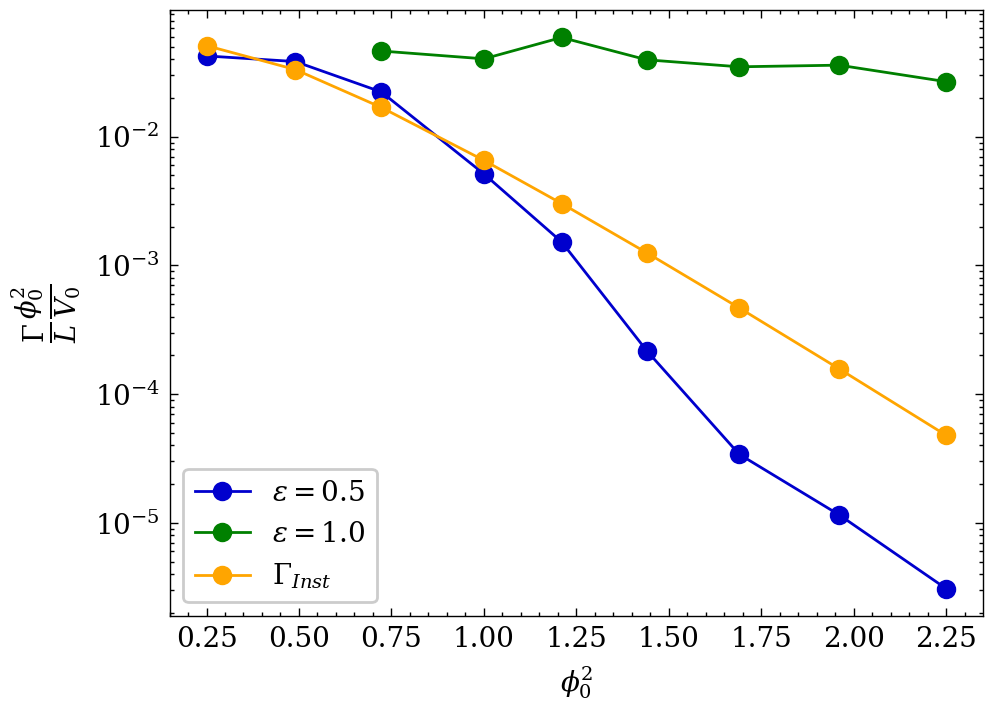}
    \caption{The nucleation rate from the instanton model and the simulations with $\epsilon = 0.5$ and $\epsilon = 1$.
    The lattice simulation has $N_x=512$, $am_f=0.06$ and integration step size $dt=0.05$.}
    \label{fig:1+1results}
\end{figure}

We first consider the system in 1+1 dimensions, exactly as in \cite{braden,hertz}. We compute the nucleation rate for different values of $\phi_0$ and for different values of the fudge factor, shown in Fig. \ref{fig:1+1results}. The number of configurations is $N=100$, which was sufficient to get convincing results. In all our figures, the (bootstrap) statistical error bars are roughly the same size as the plotting symbols. 
The instanton estimate of the quantum tunneling rate is obtained via the expression \cite{hertz} 
\be
 \frac{\Gamma}{L} = 2 m_f^2 \Big( \frac{S_B}{2 \pi} \Big) e^{-S_B} ,
\ee
Where $S_B$ refers to bounce action computed with the tool CosmoTransitions \cite{CosmoT}.
Making allowance for possible small differences in fitting procedure and numerical implementation, this reproduces the results of \cite{braden} and  \cite{hertz}, which may be summarized as follows: In 1+1 dimensions, CS simulations of a quantum vacuum-like initial ensemble produces a bubble nucleation rate of a similar order of magnitude as the quantum instanton result, at least for $\phi_0\leq 1.25$. However, this agreement is achieved from a quantum-like initial condition not with occupation numbers of 1/2, but instead 1/8, hence a fudge factor of 1/2 \cite{hertz}. In fact, tuning the fudge factor down from 1, one may achieve different levels of agreement with the instanton result at different values of $\phi_0$. The ``half'' initial condition (fudge factor 1) overestimates the quantum nucleation rate by several orders of magnitude, and has a weak dependence on the shape of the potential, $\phi_0$. 

\begin{figure}[h]
    \centering
    \includegraphics[width=12cm]{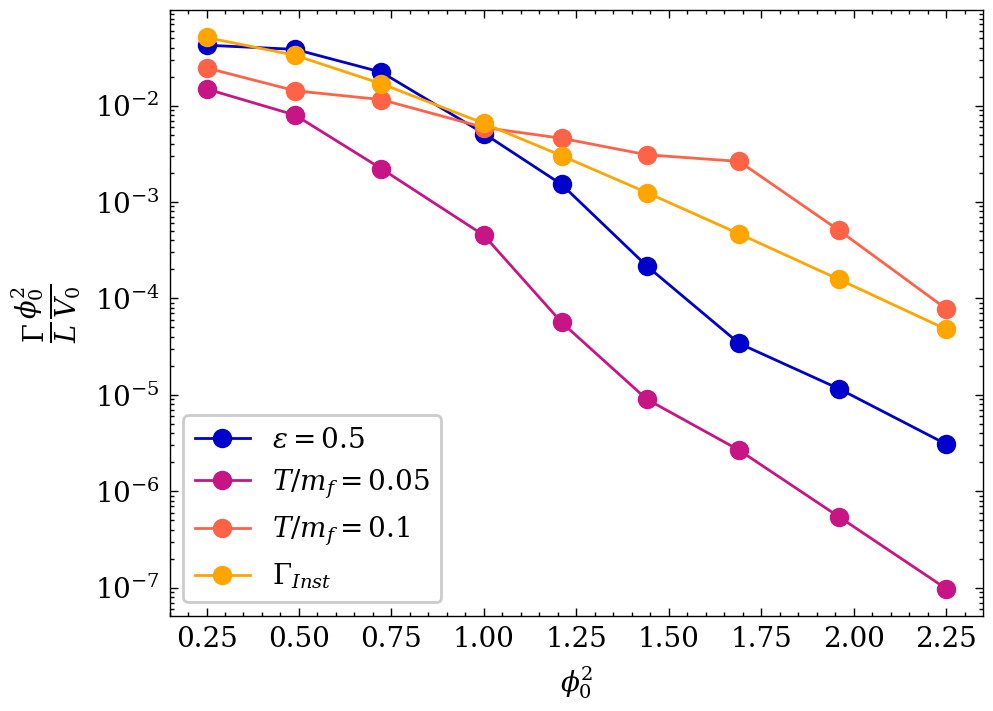}
    \caption{The thermal nucleation rate for different initial temperatures, compared to the instanton rate and the quantum-$\epsilon=0.5$ result. The cut-off is $am_f=0.3$.}
    \label{fig:1+1thermal}
\end{figure}

In Fig. \ref{fig:1+1thermal} we consider another choice of initial condition, namely the classical equilibrium mentioned above
\be
n_{\bf k}+\frac{1}{2}\rightarrow \frac{T}{\omega_k}.
\ee
While the quantum vacuum has constant occupation number for all modes, in the classical equilibrium they are suppressed in the UV. The energy density is however still divergent as the cut-off $am_f$ goes to zero. We perform the same simulation procedure as previously, but now for different values of the parameter $T$. We see that just as we did for the fudge factor $\epsilon$, we may also tune $T$ to a semi-quantitative agreement with the instanton nucleation rate, in this case $T=0.1 m_f$.

\begin{figure}[htp]
    \centering
    \includegraphics[width=12cm]{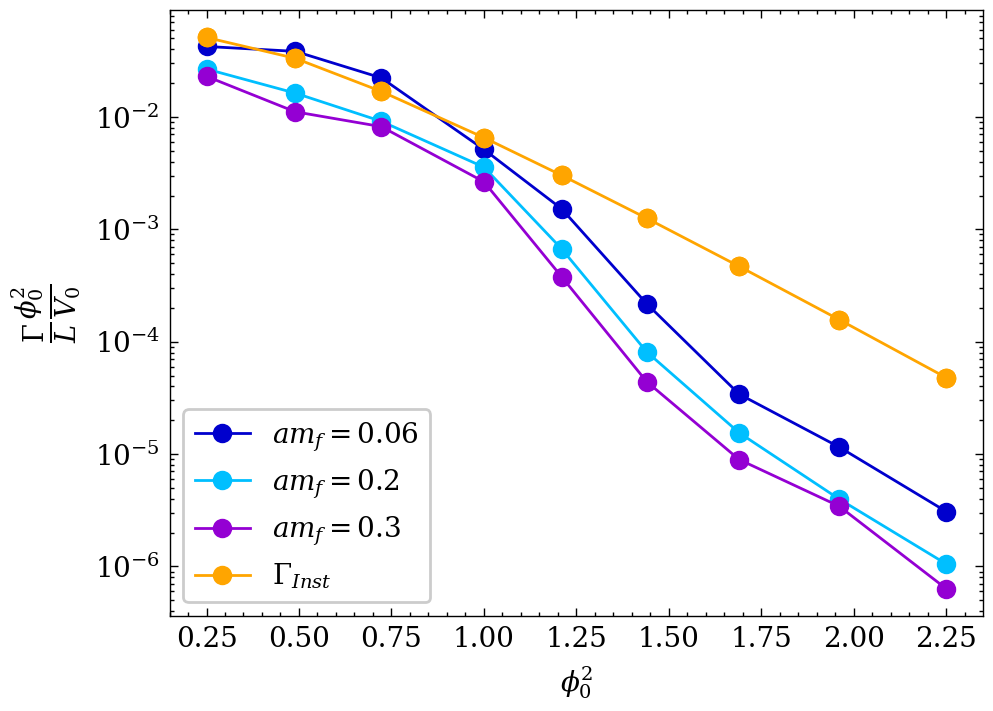}  
    \caption{The dependence of the nucleation rate on the lattice cut-off in 1+1 dimensions, for quantum-$\epsilon=0.5$ initial conditions and comparing to the instanton rate.}
    \label{fig:1+1cutoff}
\end{figure}

Since the initial conditions correspond to a divergent energy density in the continuum, it is also prudent to test the robustness of our results to changing the cut-off, in our parametrization the quantity $am_f$. The result of this procedure is shown in Fig. \ref{fig:1+1cutoff} for $\epsilon=0.5$. We see that while giving a weaker effect than varying $\epsilon$, changing the cut-off is an alternative way of tuning the rate to match the instanton rate. As might be expected, smaller $am_f$ corresponding to larger cut-off and more energy in the system leads to a larger nucleation rate. 
We may consider introducing a mass counterterm (or more generally, renormalise the potential \cite{braden2}) to counter the effect of the divergent initial condition. But because the zero point fluctuations do not stay put in classical simulations, this is difficult to achieve (see for instance \cite{Arrizabalaga:2004iw}), and does not in itself solve the problem of a divergent energy being available for tunneling. The particular potential considered here is also not readily renormalisable.

\subsection{Looking for bubbles and energy considerations in 1+1 dimensions}
\label{sec:bubblesenergy}

\begin{figure}[h]
  \centering
  \includegraphics[width=0.48\textwidth]{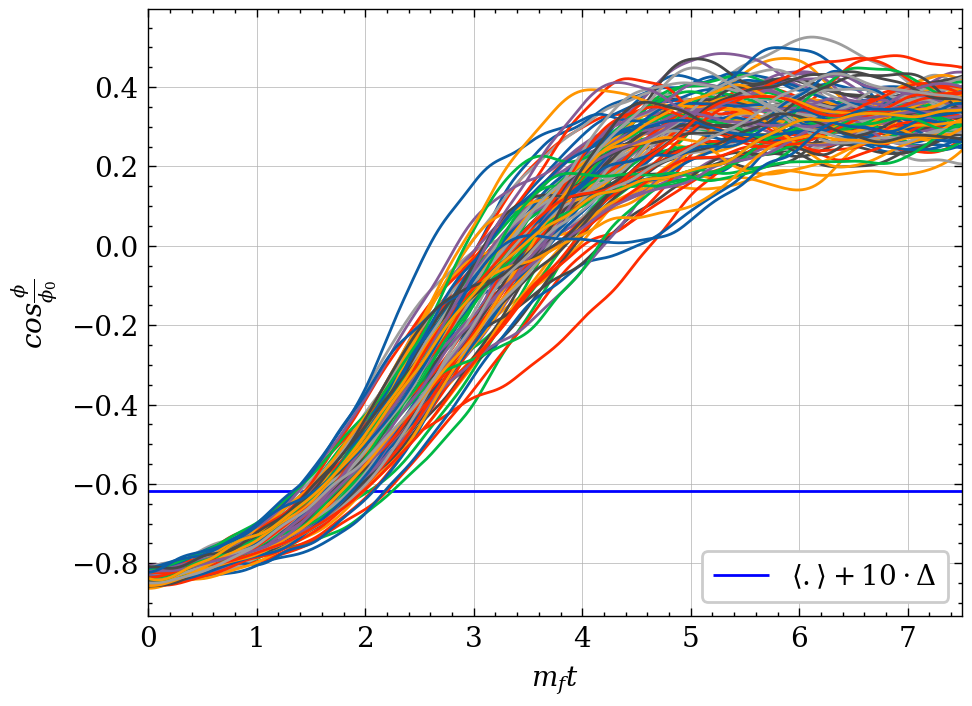}
  \includegraphics[width=0.48\textwidth]{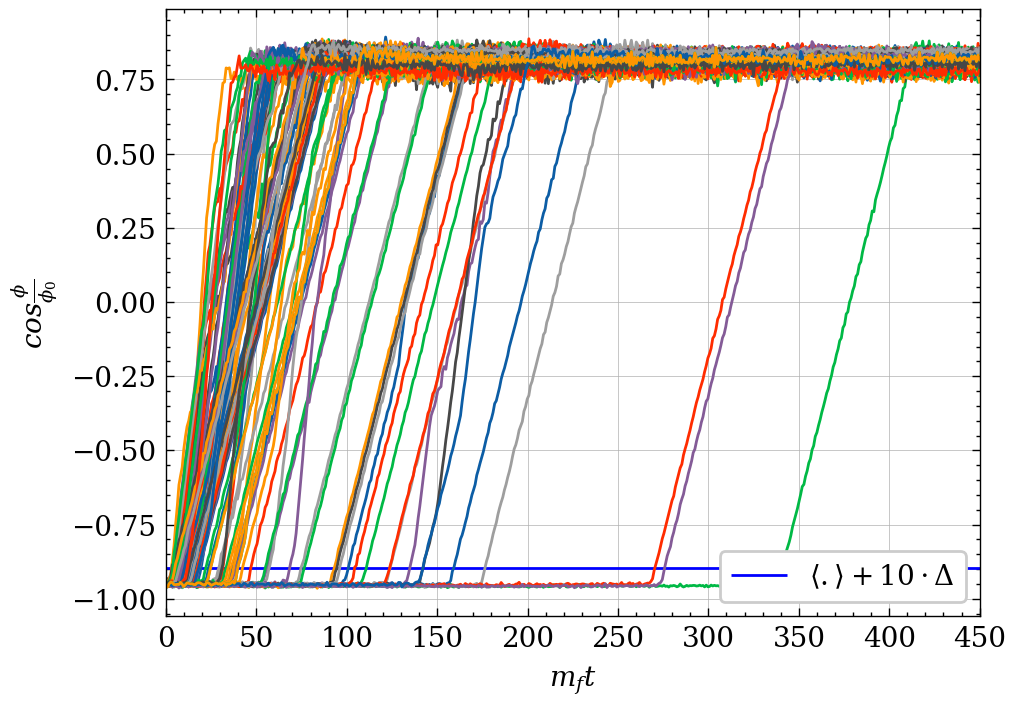}
  \includegraphics[width=0.48\textwidth]{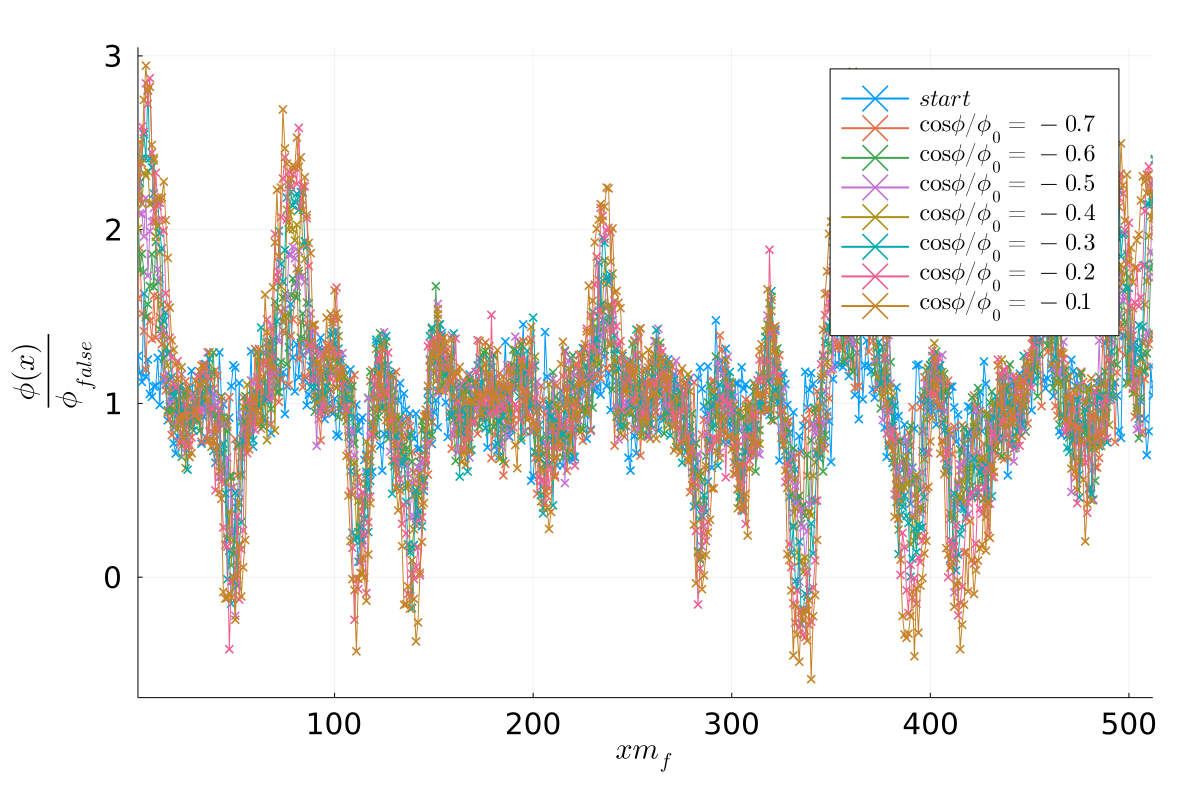}
 \includegraphics[width=0.48\textwidth]{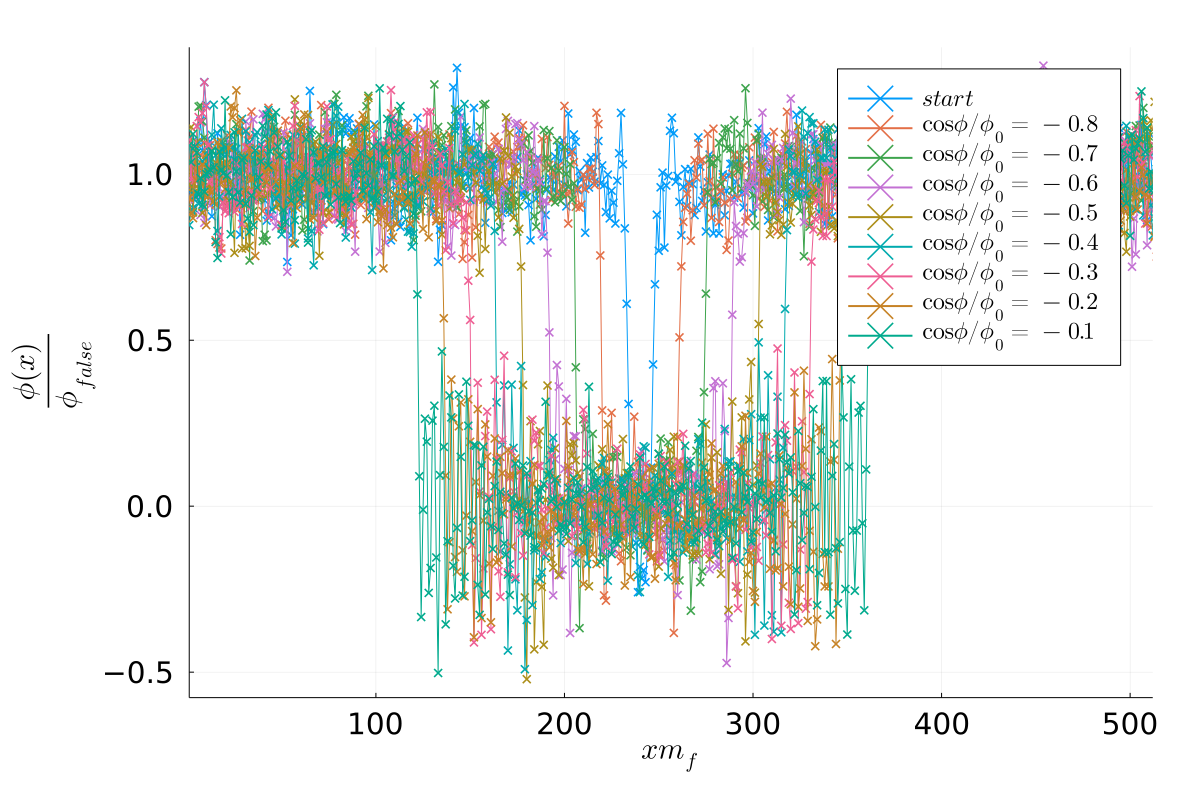}
  \caption{The time evolution of $\langle\cos(\phi/\phi_0)\rangle$ (top) for $\epsilon = 1.0$ (left) and $\epsilon = 0.5$ (right). Below are examples of configurations of one of the simulations.
  Other simulation parameters are $N_x = 512$, $dt=0.05$, $\phi_0=1.2$, $am_f=0.3$. Each ensemble consists of 100 individual configurations.}
  \label{fig:bubbleconfig}
\end{figure}

The nucleation rate for $\epsilon=1$ has a weak dependence on $\phi_0$, and similarly for $\epsilon=0.5$ for small $\phi_0$. We can begin to understand this at least qualitatively by considering the energy density of the configurations. 

The top left panel of Fig.~\ref{fig:bubbleconfig} shows the time-evolution of the 
observable $\langle\cos(\phi/\phi_0)\rangle$ for individual configurations for $\epsilon=1$ at $\phi_0=1.2$. 
The bottom left panel is one
field configuration in space at different times, labelled by the value of $\langle\cos(\phi/\phi_0)\rangle$ at that time. We see that all configurations transition through the threshold value $\simeq -0.6$ almost immediately, and that the field configurations have many nuclei and bubbles. This is an example of an initial condition with an energy density $\rho$ larger than the potential barrier $V_{\rm max}$. There is no need for the configuration to randomly organise itself into a critical bubble for the transition to take place. In contrast, the right-hand panels of Fig.~\ref{fig:bubbleconfig} show a simulation at $\epsilon=0.5$ and $\phi_0=1.2$. Here, the transitions happen as an exponential decay. Also, field configurations evolve over time from a single, initially small, bubble (light blue) to a larger and larger bubble. 

\begin{figure}[htp]
    \centering
    \includegraphics[width=12cm]{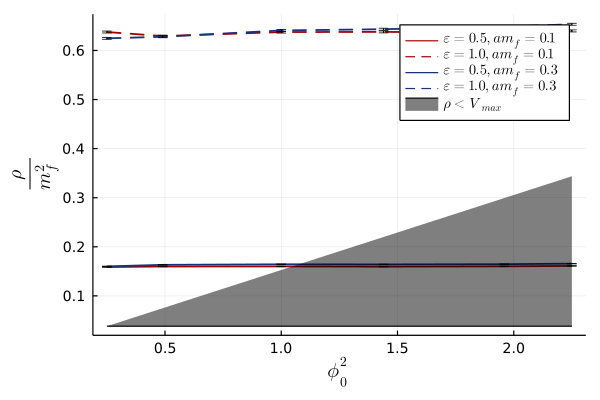}
    \caption{The dependence of energy density on $\phi_0$ and $\epsilon$. We also see a small dependence on the cut-off $am_f$.}
    \label{fig:1+1density}
\end{figure}

To make this more explicit, we compute the total energy and average energy density of the configurations.
Fig.~\ref{fig:1+1density} shows the dependence of the average energy density on $\phi_0$ and $\epsilon$.
The grey shaded region is where the average energy density
is smaller than the potential barrier, while above, the energy density is larger than the barrier. Roughly speaking, one would expect the rate of nucleation to be exponentially suppressed only in the grey region, as a critical bubble needs to emerge through a stochastic process. And one would expect the rate to be unsuppressed everywhere else. Of course, individual field configurations are inhomogeneous, multiple nuclei complicate the picture, and some out-of-equilibrium initial states may have special properties enhancing nucleation. And so the boundaries of the grey region should be considered fuzzy.  

We see that for $\epsilon=1$, we only enter the grey region far beyond the range of the figure. And that for $\epsilon=0.5$, we enter the region around $\phi_0=1.05$, corresponding roughly to where the exponential dependence on $\phi_0$ kicks in in Fig. \ref{fig:1+1results}. 

Since the energy density is approximately $\propto \epsilon^2$ and the potential barrier $\propto \phi_0^2$, the criterion for entering the grey region $\rho=V_{\rm max}$ amounts to $\phi_0\propto \epsilon$. The proportionality constant in the case depicted here happens to be $\simeq 2.05$, and so the rate for the $\epsilon=1$ initial condition of physical relevance only becomes exponentially suppressed around $\phi_0=2.1$, where the instanton rate is  $ \frac{\Gamma}{L} \frac{\phi_0^2}{V_0} =  4.7 \times 10^{-9}$. 

\begin{figure}[htp]
    \centering
    \includegraphics[width=10cm]{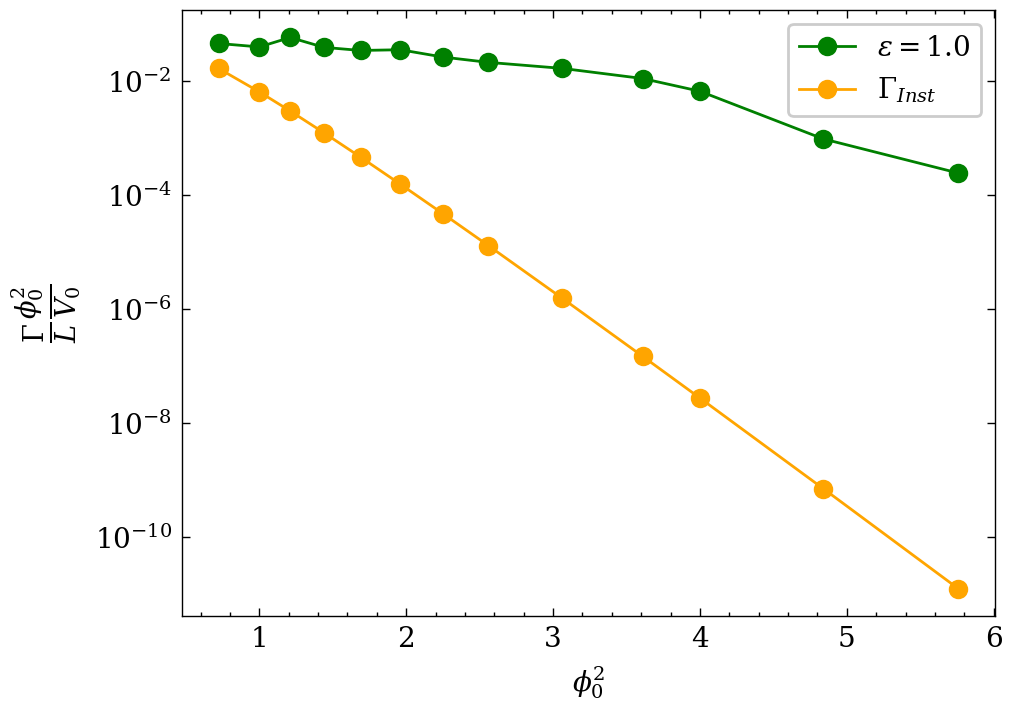}
    \caption{Tunneling rate in 1+1 dimension for larger values of  $\phi_0$. Simulations parameters are $N_x=512$, $am_f=0.06$.}
    \label{fig:1+1rateextended}
\end{figure}
In Fig. \ref{fig:1+1rateextended} we have extended the range of Fig. \ref{fig:1+1results} to include the exponentially suppressed region for $\epsilon=1$. We see again that the CS approximation overestimates the nucleation rate by several orders of magnitude.

\begin{figure}[htp]
    \centering
    \includegraphics[width=10cm]{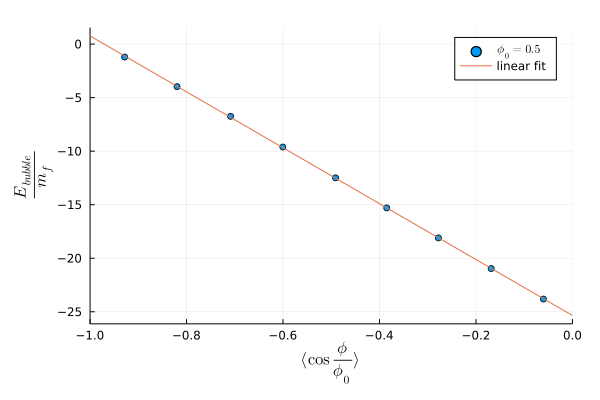}
    \caption{The energy of two bubble walls and the bubble interior versus $\langle \cos \frac{\phi}{\phi_0} \rangle$. The simulation parameters are $N_x=512$, $\phi_0=0.5$ and $\epsilon=0.17$.}
    \label{fig:Ebubb}
\end{figure}

We can attempt to compute the wall tension in 1+1 dimensions from further developing the naive model  of Sec.~\ref{sec:clastunneling}. We note that if the bubble is really in the global minimum inside the bubble ($\cos(\phi/\phi_0)=1$) and in the local minimum outside ($\cos(\phi/\phi_0)=-1$), then for a single configuration
\be
  \langle \cos \frac{\phi}{\phi_0} \rangle = \frac{4R-N_x}{N_x}, 
\ee
where $2R$ is the wall separation, and $R$ hence the radius of the bubble. We now compute numerically the energy of bubbles, where we by hand force the interior and exterior to be in the minima. Then
\be
    E_{\rm Bubble} = 2\sigma + 2 R\Delta V  = a + b \cos \frac{\phi}{\phi_0}.
\ee
We fit the parameters $a$ and $b$ for each critical bubble and relate to $\sigma $ and $\Delta V$ via
\be
\sigma = a-b; \quad \Delta V = \frac{2b}{N_x}.
\ee
In this way, an estimate for $\sigma$ can be obtained by extrapolating $E_{\rm Bubble}$ to  $\langle \cos \frac{\phi}{\phi_0} \rangle = -1$. Fig. \ref{fig:Ebubb} shows an example fit to a critical bubble obtained with $\phi_0=0.5$,  $\epsilon=0.17$.

\begin{figure}
\centering
\begin{tikzpicture}
\begin{axis}[
    xlabel={$\phi_0^2$},
    ylabel={$\frac{E_{bubble}}{m_{f}}$},
    xmin=0.25, xmax=2.25,
    ymin=0, ymax=10,
    legend pos=north west,
    ymajorgrids=true,
    grid style=dashed,
]

\addplot[
    color=blue,
    mark=square,
    ]
    coordinates {
    (0.25,1.3)(0.49,1.85)(1,3.86)(1.44,6.33)(1.96,9.22)(2.25,9.51)
    };
\end{axis}
\end{tikzpicture}
\caption{The critical bubble energy in 1+1 dimensions for different values of $\phi_0$.}
    \label{fig:Ecrit}
\end{figure}
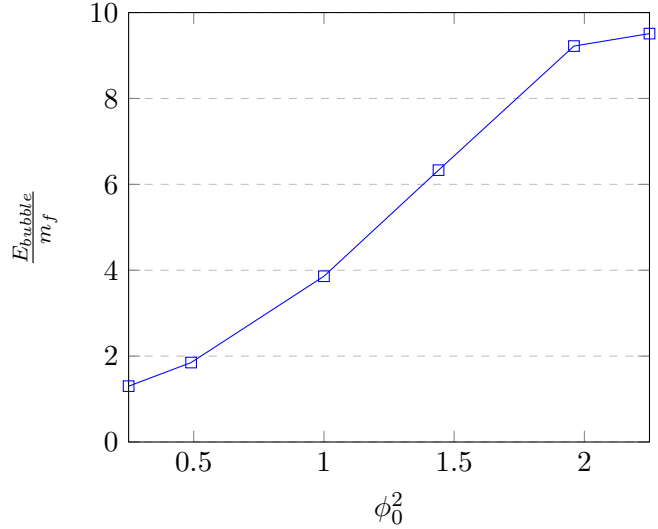

Fig.~\ref{fig:Ecrit} shows the corresponding energy of the critical bubble, $2\sigma$, for different values of $\phi_0$. This then is the minimal energy required for a configuration to classically transition to the global minimum. 
If the volume, cut-off, and $\epsilon$ is such that the energy is smaller than this value, the evolution of this interacting scalar field initially in an out-of-equilibrium state, will eventually drive the system to the classical equilibrium state in the local minimum.

\subsection{Energy depletion and thermalization in 1+1 dimensions}
\label{sec:thermalisation}

\begin{figure}[h]
  \centering
\includegraphics[width=0.7\textwidth]{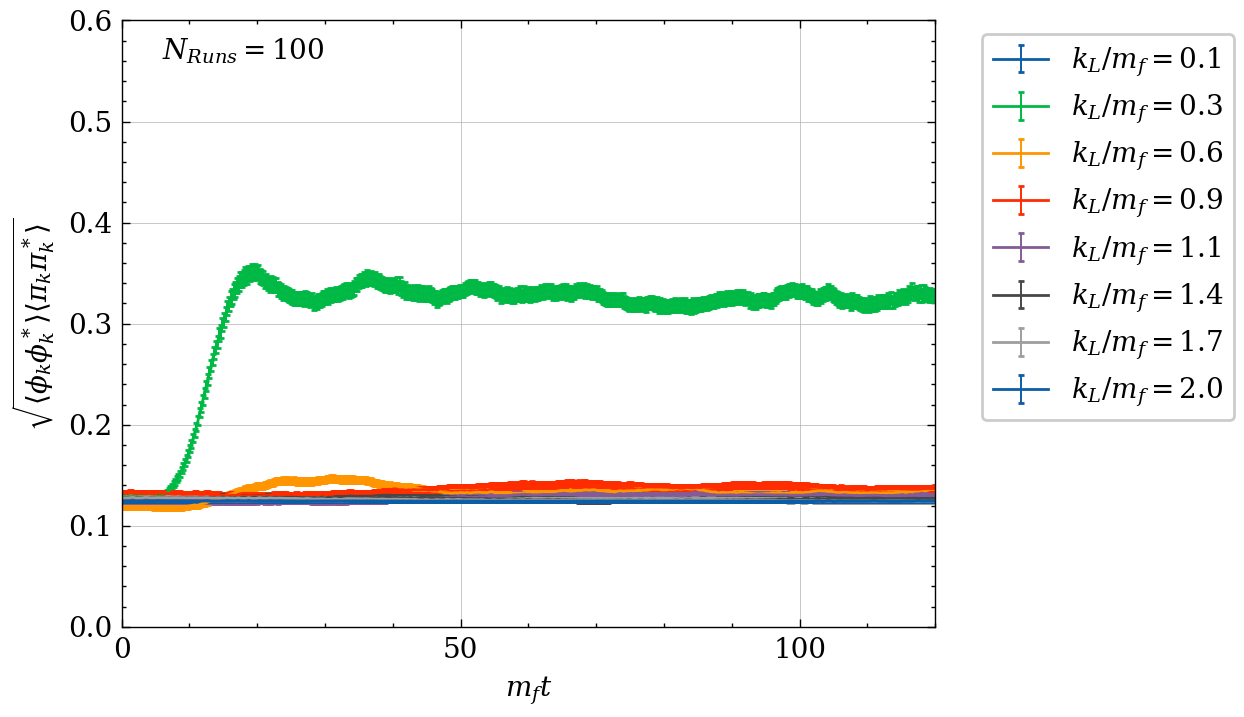}
  \caption{The evolution of the occupation numbers in 1+1 dimensions, $\epsilon=0.5$, $\phi_0=1.5$, $am_f=0.06$.}
  \label{fig:1+1particlenumbers1}
\end{figure}
 \begin{figure}[h]
  \centering
  \includegraphics[width=0.7\textwidth]{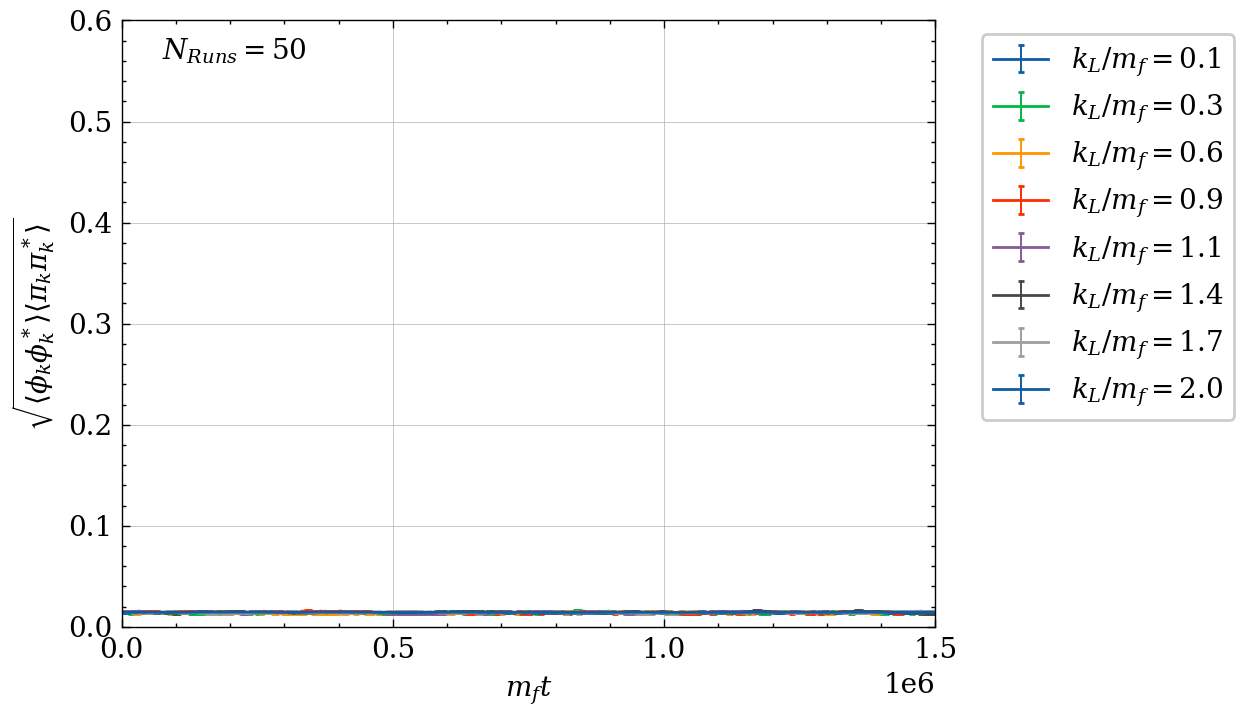}
  \caption{The evolution of the occupation numbers in 1+1 dimensions,
  $\epsilon=0.17$, $\phi_0=1.5$, $am_f=0.06$.}
  \label{fig:1+1particlenumbers2}
\end{figure}
Since the occupation numbers of the modes play an important role in the CS approximation, we also compute these through
\be
n_{\bf k}+1/2= \sqrt{\langle\pi_{\bf k}^\dagger\pi_{\bf k}\rangle\langle\phi_{\bf k}^\dagger\phi_{\bf k}\rangle}.
\label{eq:nkdef}
\ee
In Fig. \ref{fig:1+1particlenumbers1} we show the occupation numbers for a set of modes in time\footnote{The modes are collected in finite bins with several modes in each, enumerated by their central $k_L$-value.}. We see that initially, the occupation numbers (\ref{eq:nkdef}) are indeed $\epsilon^2/2$, and as the nucleation is triggered (within a time of a few in mass units), they increase as potential energy is converted into excitations. The energy is mostly deposited in IR modes.

As discussed in the preceding section, we can engineer an initial configuration with total energy less than the $E_{\rm crit}$ which will never transition. An example of this is shown in Fig. \ref{fig:1+1particlenumbers2}. For very long time, we expect the particle numbers to slowly reorganise themselves into a classical thermal spectrum. Clearly, in 1+1 dimensions, this is an extremely long time, longer than we are able to simulate. This also implies that the nucleation rates that we have computed so far indeed arise from the quantum-like initial state. It is not such, that the initial state first thermalises to the equilibrium, after which the nucleation takes place. We will return to this point when considering 2+1 dimensional simulations.


\section{Generalising to 2+1 dimensions.}
\label{sec:2+1D}


\begin{figure}[h]
    \centering
    \includegraphics[width=12cm]{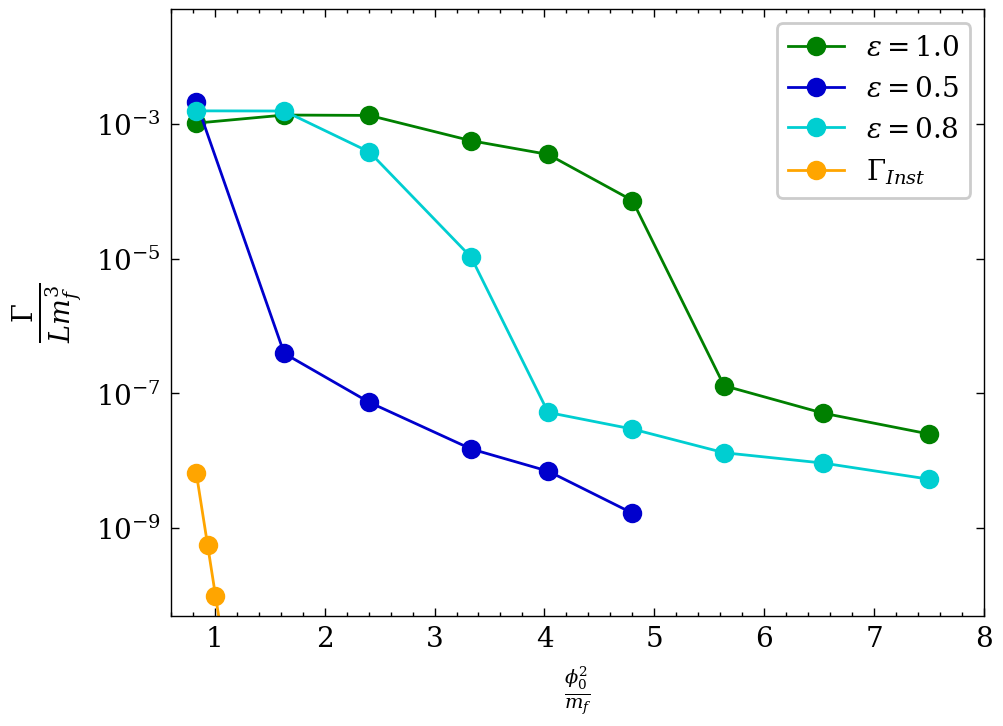}
    \caption{The nucleation rate for different values of $\phi_0$ and $\epsilon$. 
    We used $N_x^{2} = 512^2$, except for $\epsilon=0.5$ where $N_x^{2} = 128^2$ was used. We take
    $am_f = 0.3$ and $dt=0.05$.}
    \label{fig:1+2rate}
\end{figure}

We now perform 2+1 dimensional simulations completely analogously to the 1+1 dimensional case.
We discretize
the 2+1 dimensional action on a quadratic lattice of size $N_x^2$. The scale is still set by the mass $am_f$, and the other dimensionless combinations are now $a^3V_0$ and $a^{1/2}\phi_0$.
The instanton prediction is obtained by generalizing the 1+1 dimensional case, and is given by
\be
 \frac{\Gamma}{L^2} = 2 m_f^3 \Big( \frac{S_B}{2 \pi} \Big)^{3/2} e^{-S_B},
\ee
where the bounce action $S_B$ is again obtained from CosmoTransitions \cite{CosmoT}.

In Fig.~\ref{fig:1+2rate} we show the transition rates for different values
of $\phi_0$ for lattice simulations with fudge factors $\epsilon=1$, 
$\epsilon=0.8$, $\epsilon=0.5$, respectively, and again comparing to the instanton prediction.
It is clear that the CS rate even when applying a fudge factor vastly overestimates the quantum tunneling rate. Quantum bubble nucleation and false vacuum decay cannot be modelled in this way. Simulations for large values of $\phi_0$ and $\epsilon=0.5$ did not transition to a global minimum at all.
 
\begin{figure}[h]
    \centering
    \includegraphics[width=12cm]{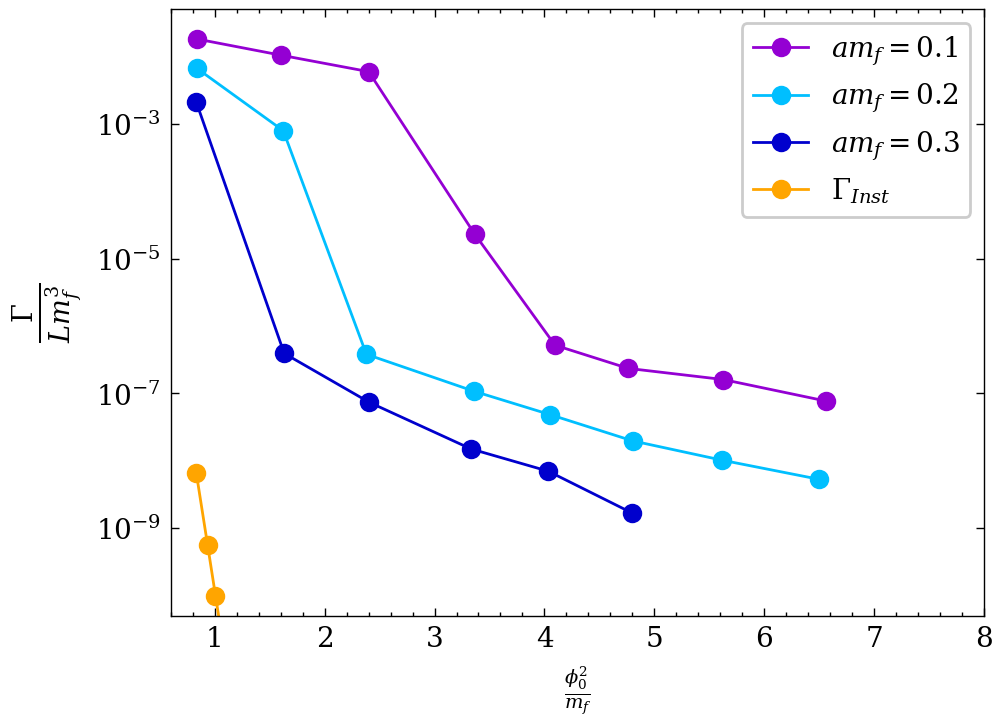}
    \caption{The cut-off dependence of the nucleation rate, again for different values $\phi_0$.}
    \label{fig:1+2ratecutoff}
\end{figure}

Fig.~\ref{fig:1+2ratecutoff} shows the nucleation rate for different values of the cut-off $am_f$ for $\epsilon=0.5$. As in 1+1 dimensions a higher cutoff (low $am_f$) results in higher rates, but the dependence is much stronger in 2+1 dimensions than it was in 1+1 dimensions. In 2 spatial dimensions, the number of UV modes grows much faster as the cut-off increases, and the initial energy density then also increases faster.

\subsection{Looking for bubbles and energy considerations in 2+1 dimensions}
\label{sec:2+1bubblesenergy}
\begin{figure}[h]
  \centering
 \includegraphics[width=0.45\textwidth]{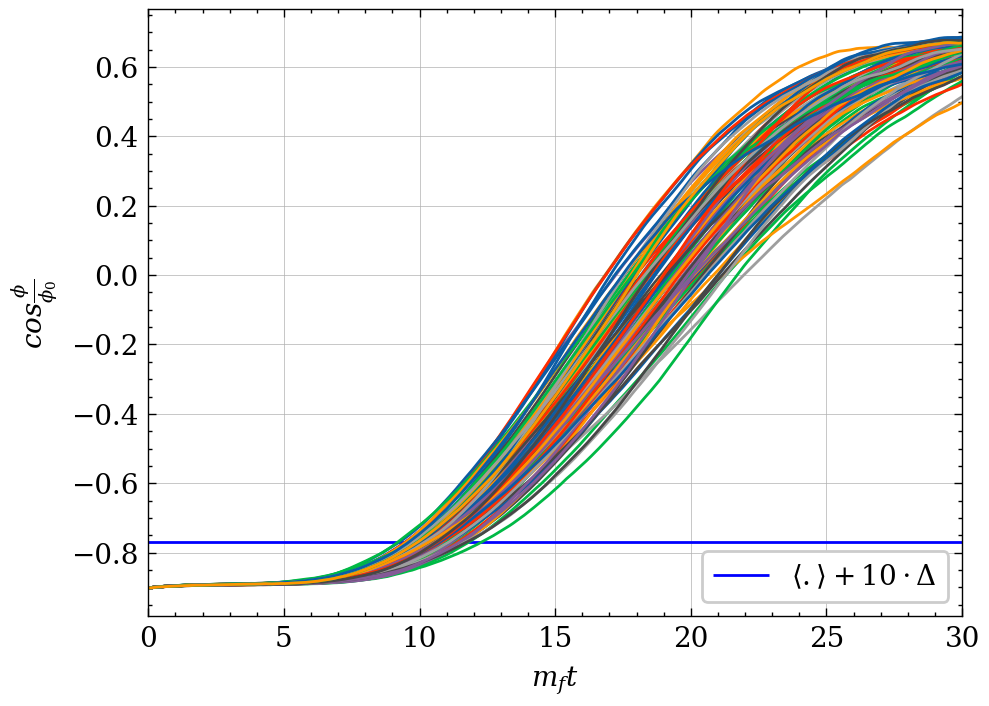}
  \includegraphics[width=0.45\textwidth]{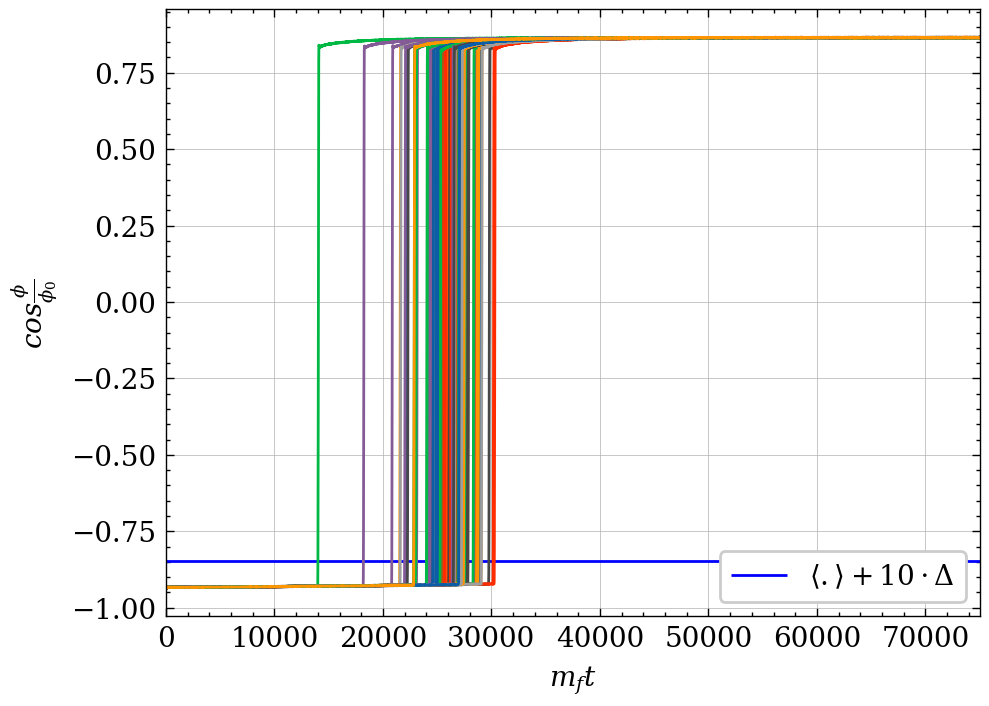}
  \includegraphics[width=0.45\textwidth]{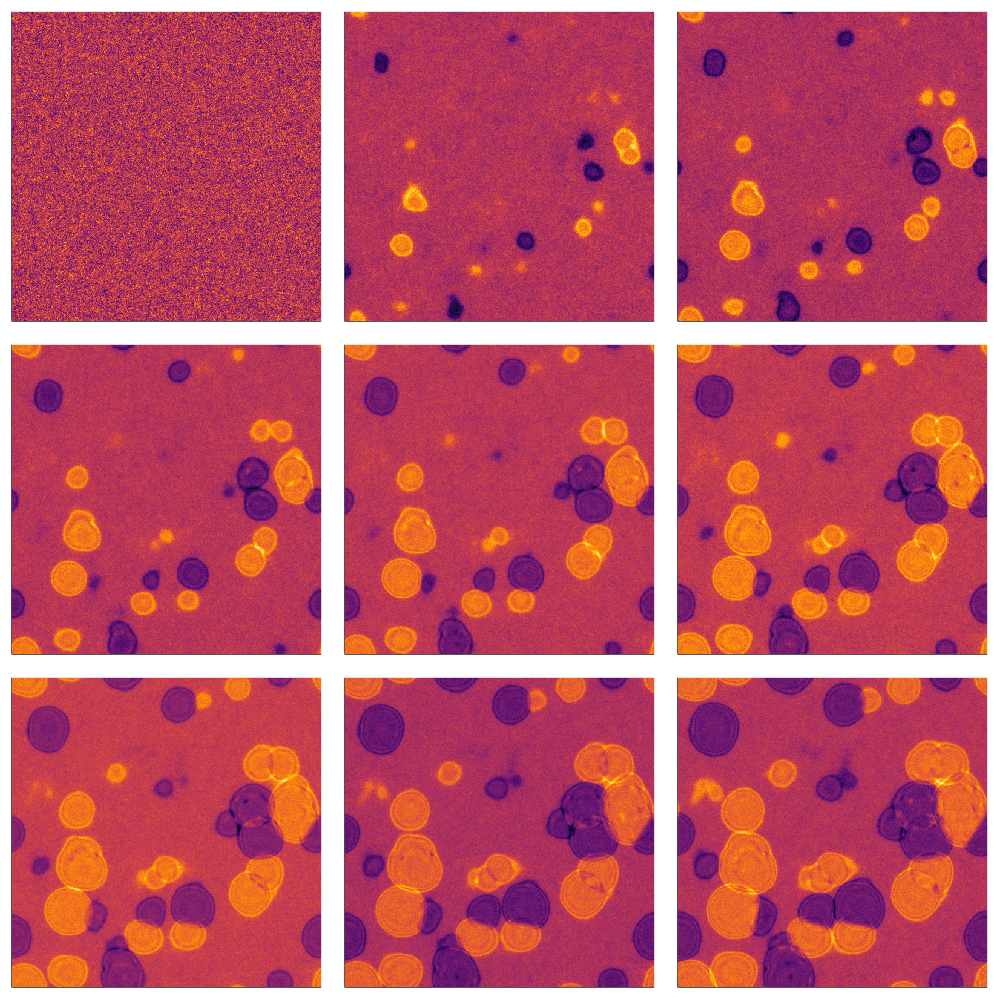}
 \includegraphics[width=0.45\textwidth]{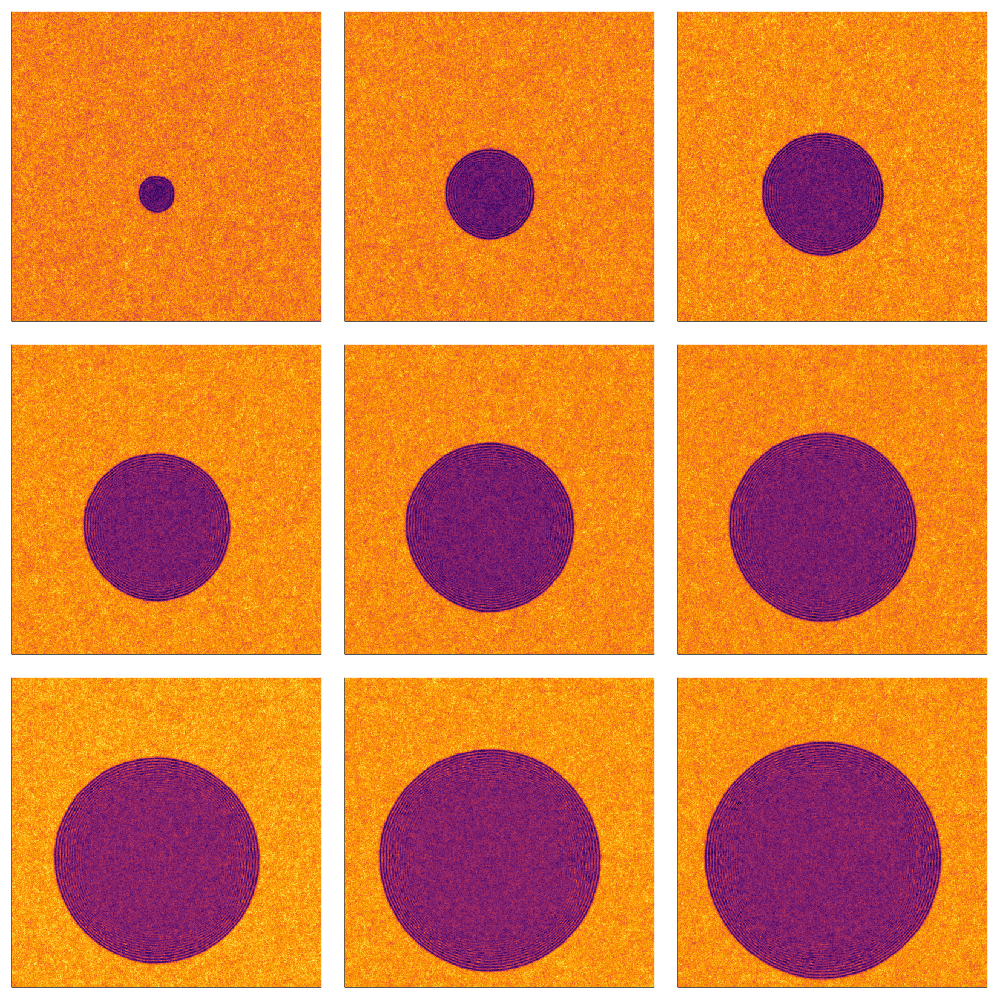}
  \caption{The time evolution of $\langle\cos(\phi/\phi_0)\rangle$ (top) for $\epsilon = 1.0$ (left) and $\epsilon = 0.5$ (right), both at $\phi_0=1.2$. Below are example configurations, with snapshots at different times in the evolution, again for $\epsilon = 1.0$ (left) and $\epsilon = 0.5$ (right).
 The simulation parameters are $N_x = 512$, $dt=0.05$, and $am_f=0.3$. Each ensemble consists of 100 individual configurations.}
  \label{fig:1+2bubbles}
\end{figure}

We will again take a closer look at the configurations close to the transition. Fig.~\ref{fig:1+2bubbles} shows two sets of simulations with high and low transition rates, respectively.
The simulations shown in the left-hand panels have $\epsilon=1.0$ and $\phi_0=1.2$ while the right-hand panels correspond to $\epsilon=0.5$ and $\phi_0=1.2$.

We observe a clear difference in that on the left-hand side, transitions happen almost immediately, and the field configurations display many small bubbles nucleating close to each other. As in 1+1 dimensions, this is a case of the energy density being larger than the potential barrier. On the right-hand side, we have an exponential decay, with just a single bubble nucleating in the entire volume. 

\begin{figure}[htp]
    \centering
    \includegraphics[width=12cm]{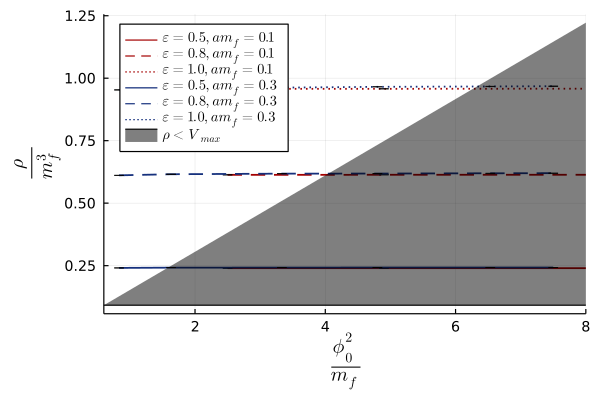}
    \caption{The initial energy density for 2+1 dimensional configurations for different values of
    $\epsilon$, $\phi_0$ and cut-off $am_f$. Configurations with $\epsilon=0.5$ can have a smaller
    average energy density than the potential barrier.}
    \label{fig:2+1density}
\end{figure}

In a similar way as in 1+1 dimensions, we compute the initial energy density and compare it to the potential barrier. This is shown in Fig.~\ref{fig:2+1density} where again the grey area corresponds to parameter combinations where the energy density is smaller than the barrier, and a bubble must be created for the transition to take place.

\begin{figure}[htp]
    \centering
    \includegraphics[width=10cm]{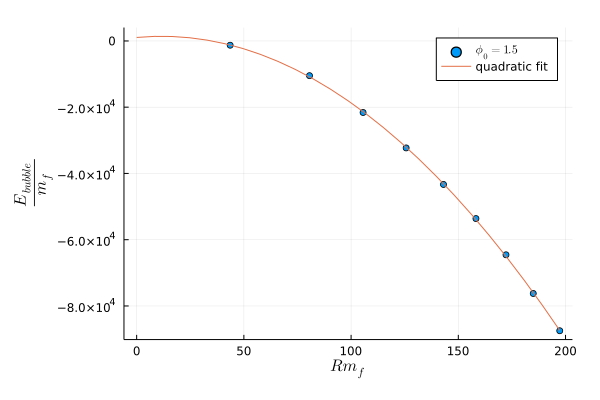}
    \caption{The energy of the bubble as a function of its radius. The simulations parameters are $N_x^2=512^2$, $\phi_0=1.5$, $am_f=0.3$ and $\epsilon=0.8$.}
    \label{fig:Ebubb2}
\end{figure}

We can again estimate the wall tension by computing the energy of 2-dimensional bubbles, but by hand fixing the inside and outside to the local and global minimum values.  
Fig.~\ref{fig:Ebubb2} shows the energy as a function of the radius of a growing bubble. As we argued in Sec.~\ref{sec:clastunneling}, the critical bubble in 2+1 and higher dimensions have a non-zero $R_{\rm crit}$, in contrast to the 1+1 dimensional case.
From the quadratic fit in the figure we can tentatively estimate the critical bubble size to be $Rm_f=15-20$. 

\subsection{Energy depletion and thermalization in 2+1 dimensions}
\label{sec:2+1thermalisation}

\begin{figure}[h]
  \centering
 \includegraphics[width=0.7\textwidth]{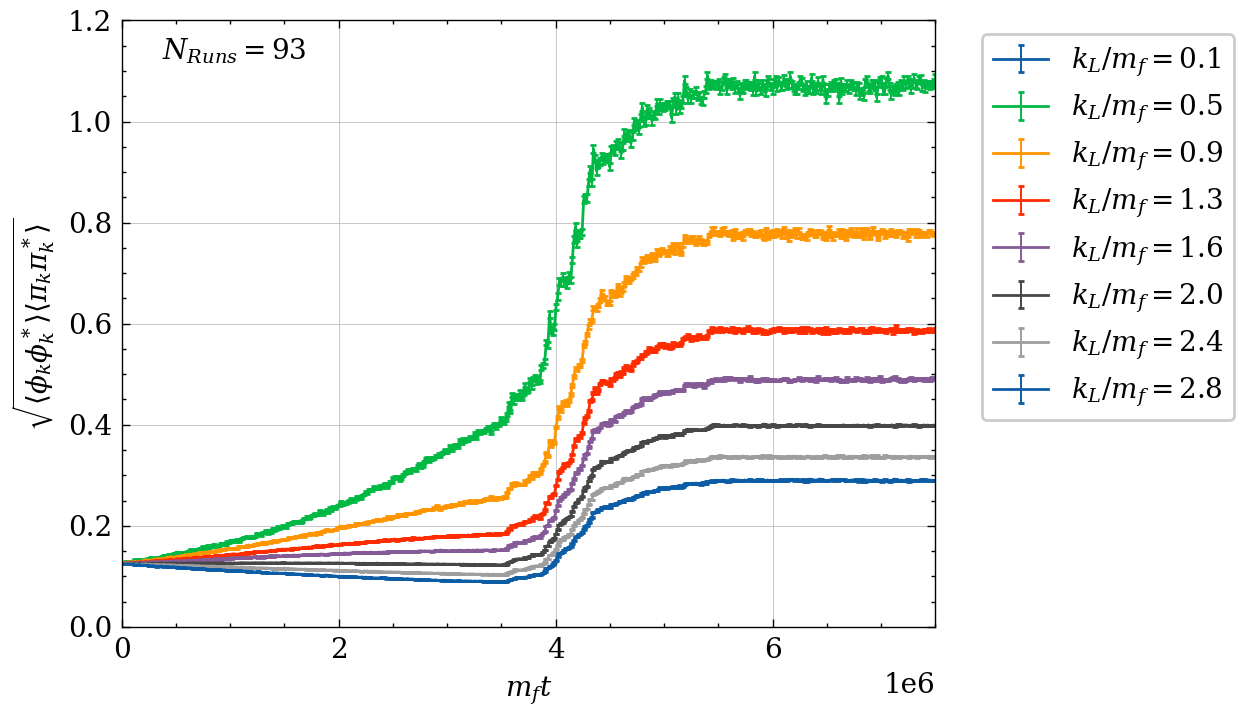}
 \includegraphics[width=0.7\textwidth]{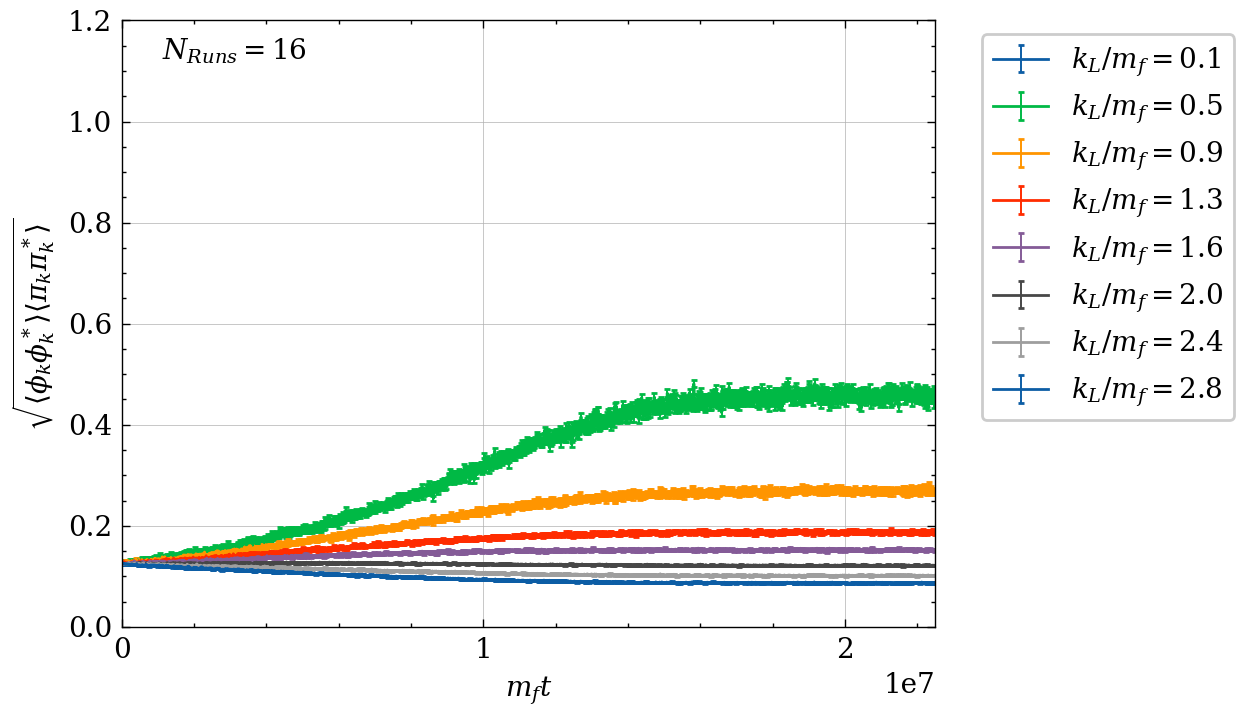}
  \caption{Time evolution of occupation numbers for simulations with $N_x=128$, $\epsilon=0.5$ and different values of $\phi_0$. Top $\phi_0=1.2$, bottom $\phi_0=1.4$}
  \label{fig:2+1nkplots}
\end{figure}

Finally, we will consider the evolution of the occupation numbers of the fields, also in 2+1 dimensions.
Fig.~\ref{fig:2+1nkplots} shows the occupation numbers as defined in (\ref{eq:nkdef}) for simulations with $\epsilon=0.5$. The top panel shows the case of $\phi_0=1.2$, where the configurations transition around $m_ft=4\times 10^6$. The bottom panel has $\phi_0=1.4$, where the configuration does not transition at all (note the time-axis extends to $2\times 10^7$).

In 1+1 dimensions, we saw that the initial quantum-like distribution is essentially unchanged up until the transition takes place. But in 2+1 dimensions, even before the transition happens the dynamics have begun redistributing the energy to approach the thermal equilibrium state. For $\phi_0=1.2$ this process does not have time to complete, but for $\phi_0=1.4$, the transition rate is so small, that the system thermalises, reaching an asymptotic state. This would not happen in the true quantum system.

It seems that in 1+1 dimensions, the time scales are such that classical nucleation is always much faster than thermalization. While in 2+1 dimensions, kinetic equilibration is often well underway by the time the transition happens. This ordering of time scales is dependent on the potential (the strength of self-interactions), the initial condition ($\epsilon$, say) and the cut-off ($am_f$). 


\section{Conclusions}
\label{sec:Conclusions}

Motivated by the intriguing possibility proposed in \cite{braden}, that classical-statistical simulations could have something to say about quantum vacuum decay, we have investigated such simulations, both in 1+1 and 2+1 dimensions. The conclusion is disappointing, although perhaps not wholly unexpected.

As also demonstrated in \cite{hertz}, the reported approximate agreement between the instanton calculation and the CS simulations is there, but as we have seen it arises through arbitrarily adjusting the parameters of the initial conditions (the amplitude $\epsilon$, cut-off $am_f$), and is also not specific to the quantum-like state with equal occupation numbers in all modes (thermal initial conditions work just as well, when tuning $T$). In fact, the actual, $\epsilon=1$, ``half'' initial condition intended to be mimic the zero-point fluctuations of the false vacuum produces a nucleation rate several orders of magnitude larger than the instanton nucleation rate, also in the range of $\phi_0$, where the energy density is smaller than the barrier. In addition, obtaining even approximate agreement between CS simulations and the instanton result is specific to 1+1 dimensions. In 2+1 dimensions, the CS simulations consistently overestimate the nucleation rate by many orders of magnitude. We also attempted simulations along the same lines for the physically relevant case of 3+1 dimensions, but the nucleation rate is then far below our numerical reach, and advanced Monte-Carlo techniques are likely required to compute also the classical rate \cite{Moore:2000jw,Moore:2001vf,Gould:2022ran}.  

As mentioned in the introduction, the CS-approximation may be derived directly as a limit of the full real-time path integral \cite{Mou:2019gyl}. It is only a good approximation for interacting quantum evolution for large occupation numbers, and even then only when computing ``classical'' observables. Quantum vacuum decay is both inherently quantum and by construction has an initial condition with occupation numbers $\ll 1$. Such initial conditions can only reliably be simulated in the CS approximation for very small coupling, when the evolution equations are (approximately) linear. But we have seen that even the proposed ``half'' initial condition probes the non-linear regions of the potential considered here (Fig. \ref{fig:Potdist}). 

We conclude that computing quantum tunneling rates in field theory beyond \cite{colman1} remains a difficult task, which cannot be simulated using classical dynamics of an ensemble of configurations. It likely requires non-perturbative numerical methods at the level of the path integral, known to be challenging for real-time systems out of equilibrium (although see \cite{Berges:2000ur,Berges:2004yj,Arrizabalaga:2004iw, Arrizabalaga:2005tf} and \cite{Mou:2019gyl}). Fortunately, in almost all cases, phase transitions involve non-vacuum initial states, for which the quantum rate is insignificant compared to the classical nucleation rate. And classical nucleation rates may be computed using CS simulations or stochastic evolution in effective theories \cite{Moore:2000jw,Moore:2001vf}.

\section*{Acknowledgement}
We thank Paul Saffin, Peter Millington, Zong-Gang Mou and Alexander Rothkopf for collaboration on related topics as well as useful discussions and comments on the present manuscript. 

\bibliography{biblio.bib}

\begin{thebibliography}{10}

\bibitem{braden}
Jonathan Braden, Matthew~C. Johnson, Hiranya~V. Peiris, Andrew Pontzen, and
  Silke Weinfurtner.
\newblock New semiclassical picture of vacuum decay.
\newblock {\em Phys. Rev. Lett.}, 123:031601, Jul 2019.

\bibitem{Hertzberg:2016tal}
Mark~P. Hertzberg.
\newblock {Quantum and Classical Behavior in Interacting Bosonic Systems}.
\newblock {\em JCAP}, 11:037, 2016.

\bibitem{Aarts:1997kp}
Gert Aarts and Jan Smit.
\newblock {Classical approximation for time dependent quantum field theory:
  Diagrammatic analysis for hot scalar fields}.
\newblock {\em Nucl. Phys. B}, 511:451--478, 1998.

\bibitem{Aarts:2001yn}
Gert Aarts and Juergen Berges.
\newblock {Classical aspects of quantum fields far from equilibrium}.
\newblock {\em Phys. Rev. Lett.}, 88:041603, 2002.

\bibitem{Rajantie:2006gy}
Arttu Rajantie and Anders Tranberg.
\newblock {Looking for defects in the 2PI correlator}.
\newblock {\em JHEP}, 11:020, 2006.

\bibitem{Arrizabalaga:2005tf}
Alejandro Arrizabalaga, Jan Smit, and Anders Tranberg.
\newblock {Equilibration in phi**4 theory in 3+1 dimensions}.
\newblock {\em Phys. Rev. D}, 72:025014, 2005.

\bibitem{Mou:2019gyl}
Zong-Gang Mou, Paul~M. Saffin, and Anders Tranberg.
\newblock {Quantum tunnelling, real-time dynamics and Picard-Lefschetz
  thimbles}.
\newblock {\em JHEP}, 11:135, 2019.

\bibitem{Millington:2020vkg}
Peter Millington, Zong-Gang Mou, Paul~M. Saffin, and Anders Tranberg.
\newblock {Statistics on Lefschetz thimbles: Bell/Leggett-Garg inequalities and
  the classical-statistical approximation}.
\newblock {\em JHEP}, 03:077, 2021.

\bibitem{hertz}
Mark~P. Hertzberg, Fabrizio Rompineve, and Neil Shah.
\newblock Quantitative analysis of the stochastic approach to quantum
  tunneling.
\newblock {\em Phys. Rev. D}, 102:076003, Oct 2020.

\bibitem{Moore:1999fs}
Guy~D. Moore and Kari Rummukainen.
\newblock {Classical sphaleron rate on fine lattices}.
\newblock {\em Phys. Rev. D}, 61:105008, 2000.

\bibitem{Berges:2007re}
Juergen Berges, Sebastian Scheffler, and Denes Sexty.
\newblock {Bottom-up isotropization in classical-statistical lattice gauge
  theory}.
\newblock {\em Phys. Rev. D}, 77:034504, 2008.

\bibitem{Rajantie:2010tb}
Arttu Rajantie and Anders Tranberg.
\newblock {Counting Defects with the Two-Point Correlator}.
\newblock {\em JHEP}, 08:086, 2010.

\bibitem{DOnofrio:2014rug}
Michela D'Onofrio, Kari Rummukainen, and Anders Tranberg.
\newblock {Sphaleron Rate in the Minimal Standard Model}.
\newblock {\em Phys. Rev. Lett.}, 113(14):141602, 2014.

\bibitem{Rajantie:2000nj}
A.~Rajantie, P.~M. Saffin, and Edmund~J. Copeland.
\newblock {Electroweak preheating on a lattice}.
\newblock {\em Phys. Rev. D}, 63:123512, 2001.

\bibitem{Greene:1997fu}
Patrick~B. Greene, Lev Kofman, Andrei~D. Linde, and Alexei~A. Starobinsky.
\newblock {Structure of resonance in preheating after inflation}.
\newblock {\em Phys. Rev. D}, 56:6175--6192, 1997.

\bibitem{Kofman:1997yn}
Lev Kofman, Andrei~D. Linde, and Alexei~A. Starobinsky.
\newblock {Towards the theory of reheating after inflation}.
\newblock {\em Phys. Rev. D}, 56:3258--3295, 1997.

\bibitem{Felder:2000hj}
Gary~N. Felder, Juan Garcia-Bellido, Patrick~B. Greene, Lev Kofman, Andrei~D.
  Linde, and Igor Tkachev.
\newblock {Dynamics of symmetry breaking and tachyonic preheating}.
\newblock {\em Phys. Rev. Lett.}, 87:011601, 2001.

\bibitem{Bodeker:2007fw}
Dietrich Bodeker and Kari Rummukainen.
\newblock {Non-abelian plasma instabilities for strong anisotropy}.
\newblock {\em JHEP}, 07:022, 2007.

\bibitem{Rebhan:2004ur}
Anton Rebhan, Paul Romatschke, and Michael Strickland.
\newblock {Hard-loop dynamics of non-Abelian plasma instabilities}.
\newblock {\em Phys. Rev. Lett.}, 94:102303, 2005.

\bibitem{Tranberg:2003gi}
Anders Tranberg and Jan Smit.
\newblock {Baryon asymmetry from electroweak tachyonic preheating}.
\newblock {\em JHEP}, 11:016, 2003.

\bibitem{Garcia-Bellido:1999xos}
Juan Garcia-Bellido, Dmitri~Yu. Grigoriev, Alexander Kusenko, and Mikhail~E.
  Shaposhnikov.
\newblock {Nonequilibrium electroweak baryogenesis from preheating after
  inflation}.
\newblock {\em Phys. Rev. D}, 60:123504, 1999.

\bibitem{Garcia-Bellido:2002fsq}
Juan Garcia-Bellido, Margarita Garcia~Perez, and Antonio Gonzalez-Arroyo.
\newblock {Symmetry breaking and false vacuum decay after hybrid inflation}.
\newblock {\em Phys. Rev. D}, 67:103501, 2003.

\bibitem{Smit:2002yg}
Jan Smit and Anders Tranberg.
\newblock {Chern-Simons number asymmetry from CP violation at electroweak
  tachyonic preheating}.
\newblock {\em JHEP}, 12:020, 2002.

\bibitem{Arrizabalaga:2004iw}
Alejandro Arrizabalaga, Jan Smit, and Anders Tranberg.
\newblock {Tachyonic preheating using 2PI-1/N dynamics and the classical
  approximation}.
\newblock {\em JHEP}, 10:017, 2004.

\bibitem{Mukhanov:1990me}
Viatcheslav~F. Mukhanov, H.~A. Feldman, and Robert~H. Brandenberger.
\newblock {Theory of cosmological perturbations. Part 1. Classical
  perturbations. Part 2. Quantum theory of perturbations. Part 3. Extensions}.
\newblock {\em Phys. Rept.}, 215:203--333, 1992.

\bibitem{Amin:2014eta}
Mustafa~A. Amin, Mark~P. Hertzberg, David~I. Kaiser, and Johanna Karouby.
\newblock {Nonperturbative Dynamics Of Reheating After Inflation: A Review}.
\newblock {\em Int. J. Mod. Phys. D}, 24:1530003, 2014.

\bibitem{colman1}
Sidney Coleman.
\newblock Fate of the false vacuum: Semiclassical theory.
\newblock {\em Phys. Rev. D}, 15:2929--2936, May 1977.

\bibitem{Moore:2000jw}
Guy~D. Moore and Kari Rummukainen.
\newblock {Electroweak bubble nucleation, nonperturbatively}.
\newblock {\em Phys. Rev. D}, 63:045002, 2001.

\bibitem{Moore:2001vf}
Guy~D. Moore, Kari Rummukainen, and Anders Tranberg.
\newblock {Nonperturbative computation of the bubble nucleation rate in the
  cubic anisotropy model}.
\newblock {\em JHEP}, 04:017, 2001.

\bibitem{Gould:2022ran}
Oliver Gould, Sinan G\"uyer, and Kari Rummukainen.
\newblock {First-order electroweak phase transitions: a nonperturbative
  update}.
\newblock 5 2022.

\bibitem{CosmoT}
Carroll~L. Wainwright.
\newblock Cosmotransitions: Computing cosmological phase transition
  temperatures and bubble profiles with multiple fields.
\newblock {\em Computer Physics Communications}, 183(9):2006--2013, 2012.

\bibitem{braden2}
Jonathan Braden, Matthew~C. Johnson, Hiranya~V. Peiris, Andrew Pontzen, and
  Silke Weinfurtner.
\newblock {Mass Renormalization in Lattice Simulations of False Vacuum Decay}.
\newblock 4 2022.

\bibitem{Berges:2000ur}
Juergen Berges and Jurgen Cox.
\newblock {Thermalization of quantum fields from time reversal invariant
  evolution equations}.
\newblock {\em Phys. Lett. B}, 517:369--374, 2001.

\bibitem{Berges:2004yj}
Juergen Berges.
\newblock {Introduction to nonequilibrium quantum field theory}.
\newblock {\em AIP Conf. Proc.}, 739(1):3--62, 2004.

\end{thebibliography}
\bibliographystyle{unsrt}

\end{document}